\documentclass[reprint,
superscriptaddress,
amsmath,
amssymb,
aps,
prb,
floatfix,
english
]{revtex4-2}
\usepackage{graphicx}
\usepackage{color}
\usepackage{braket}
\usepackage{amsfonts}
\usepackage{MnSymbol}
\usepackage{bm}
\usepackage{bibunits}
\usepackage[unicode=true, colorlinks=true, citecolor={blue!80!black}, urlcolor={blue!50!black}, linkcolor = {blue!80!black}]{hyperref}
\defaultbibliography{references}
\defaultbibliographystyle{apsrev4-2}

\usepackage[dvipsnames]{xcolor}
\usepackage{mathrsfs}
\usepackage{mathtools}
\usepackage{subfigure}

\usepackage{bbm}
\usepackage[dvipsnames]{xcolor}

\date{\today}                  

\begin{document}

\title{Theory of three-terminal Andreev spin qubits}

\author{Kiryl Piasotski}
\email[Email: ]{kiryl.piasotski@kit.edu}
\affiliation{Institut f\"ur Theorie der Kondensierten Materie (TKM), Karlsruher Institut f\"ur Technologie, 76131 Karlsruhe, Germany}
\affiliation{Institut f\"ur Quanten Materialien und Technologien (IQMT), Karlsruher Institut f\"ur Technologie, 76021 Karlsruhe, Germany}

\author{Aleksandr Svetogorov}
\affiliation{Universit\"at Konstanz, 78457 Konstanz, Germany}

\author{Wolfgang Belzig}
\affiliation{Universit\"at Konstanz, 78457 Konstanz, Germany}

\author{Mikhail Pletyukhov}
\affiliation{Institut für Theorie der Statistischen Physik, 
RWTH Aachen, 52074 Aachen, Germany}

\begin{abstract}

In this paper, we introduce a concise theoretical framework for the equilibrium three-terminal Josephson effect in spin-orbit-interacting systems, inspired by recent experiments on an InAs/Al heterostructure [Phys. Rev. X {\bf 14}, 031024 (2024)]. We develop an analytical model to capture the essential low-energy physics of the system and examine its potential as an Andreev spin qubit, while also reconciling some findings of Ref. [Phys. Rev. B {\bf 90}, 155450 (2014)]. Our analysis of the transitions between the Andreev levels in the junction shows that, in an idealized scenario, the transition between the lowest pair of pseudo-spin-split Andreev levels is blocked by pseudo-spin conservation. We demonstrate that to operate the system as an Andreev spin qubit, leveraging the significant spin splitting observed experimentally, additional ingredients such as external magnetic filed or magnetic impurities are required. Finally, we apply our model to investigate the coupling between two such qubits, mediated by supercurrent.
\end{abstract}

\maketitle

\section{Introduction}

Andreev spin qubits (ASQs) have been proposed in a theoretical work~\cite{Nazarov2003} as a way to combine electron spins in localized states that can maintain coherence on long time scales~\cite{Dzurak2014,Dzurak2024} and strong long-range coupling via Cooper pairs condensate. Coupling between an electron spin and supercurrent is achieved by spin-orbit interactions (SOI)~\cite{Beenakker2008,Nazarov2010,Avignon2012,Yeyati2017,Tosi2019,Devoret2020,Heck2023}, such hybridized spin is usually refereed to as pseudo-spin. The first ASQ has recently been developed~\cite{Devoret2021} in a semiconducting nanowire, where SOI lift the degeneracy of two states of a quasiparticle localized in an Andreev bound state (ABS) between two superconducting leads. Further experiments were performed in a semiconducting quantum dot (QD), where a combination of SOI and magnetic field allowed for a visible splitting as well as a long-range coupling between two ASQs~\cite{Andersen2023,Andersen2024}. Nevertheless, to speed up two-qubit gates, stronger spin splitting is required. Therefore, we suggest an improved design of an ASQ -- a three-terminal Andreev spin qubit (TASQ) -- that demonstrates enhanced spin splitting of the ABS~\cite{Nichele2024}.

In our work, we harness a powerful Green's function formalism~\cite{Alvarado2022,zazunov2018josephson,PhysRevB.101.094511, PhysRevB.101.115405, PhysRevB.100.081106, PhysRevB.96.155103, PhysRevB.96.024516, PhysRevB.95.235143, PhysRevB.94.014502, PhysRevB.92.085126, PhysRevB.80.014425, PhysRevB.106.165405} developed further in a recent theoretical work~\cite{Piasotski2024} for quasi-one-dimensional channels between superconducting leads. Such an approach allows us to establish the non-trivial properties of a three-lead setup in comparison to a Josephson junction with two leads. We further develop the ideas put forth in a theoretical work~\cite{Akhmerov2014}, such as the enhanced phase-dependent spin splitting of the bound states and changes in the ground state fermion parity. 

Our model captures features observed in a recent experiment~\cite{Nichele2024} and allows us to propose an optimal regime for the qubit operation. Our findings combined with experimental evidence~\cite{Nichele2024} suggest that the TASQ is a feasible device that overcomes one of the main disadvantages of a two-terminal ASQ -- the weak splitting of ABSs. The higher connectivity of the TASQ as a circuit node enables a wider range of multi-qubit architectures, leveraging the robust long-distance supercurrent-mediated inter-qubit coupling.

\begin{figure}[b!]
              \includegraphics[width=0.43\textwidth]{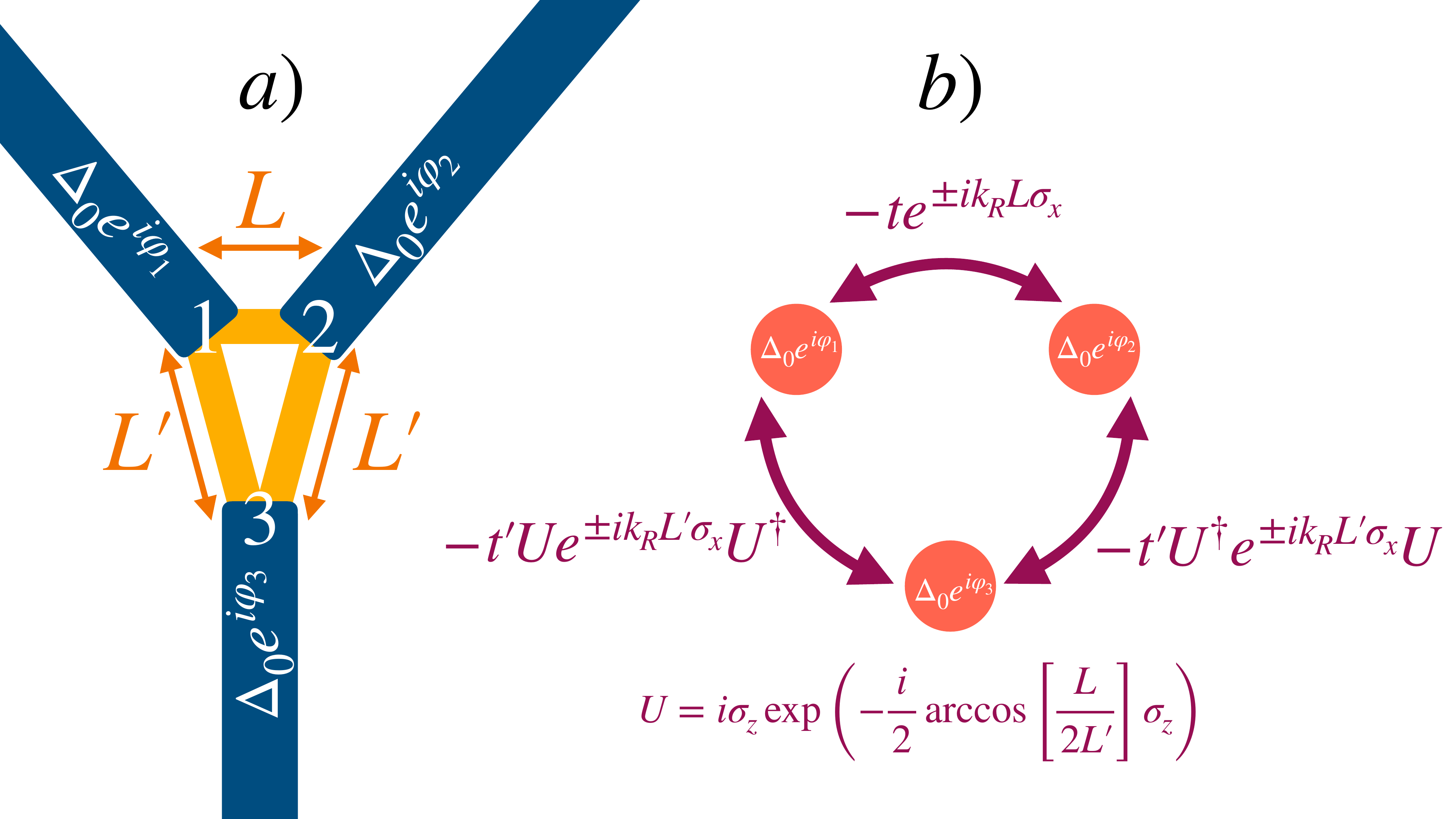}
                \caption{Sketch of the effective model's configuration. As illustrated in Panel a), the effective model replaces the semiconducting plate (where the superconducting electrodes connect) with three quantum wires linking the contacts (see SM \cite{SM} for details). By further integrating the extended components of the system and projecting onto the Fermi surface, we can establish (see SM \cite{SM}) an effective three-site model based on the contact points of the original setup, as described by Eq. \eqref{H_eff_main} and depicted schematically in Panel b). Additionally, the spin-momentum locking from the original model results in rotations of spin-dependent hopping amplitudes as one moves from one contact to another, as is further indicated schematically in  Panel b).}
               \label{fig: model_sketch}
\end{figure}

\section{Quantum model and its spectrum}\label{sec:model}

In this section, we present a simple low-energy description of three-terminal Josephson junctions in two-dimensional electron gas with spin-orbit interaction. We begin by noting that the dominant portion of the current flowing between the superconductors through the semiconducting plate is transmitted via one-dimensional ballistic channels connecting the contact points. Further leveraging the techniques of Ref.~\cite{Piasotski2024} to integrate these channels out, enables the construction of an effective low-energy model in the space of the three contact points, akin to the empirical quantum dot model proposed in Ref. \cite{Heck2023}.

As is earlier anticipated, to gain a basic understanding, we replace the semiconducting plate connecting the three superconducting electrodes with a triangle made up of one-dimensional ballistic channels that link the contact points of the electrodes (see the left panel of Fig. \ref{fig: model_sketch}). It is crucial to note that, as the SOI in the original semiconducting plate is assumed to be constant throughout the sample \footnote{In the semiconducting part of the system,  the Rashba Hamiltonian, traditionally written in the form  $H_R = \alpha_R (-k_x \sigma_y + k_y \sigma_x)$, describes the spin-orbit coupling in the two-dimensional electron gas. Our convention, however, is to represent it as $\alpha_R \, \vec{k} \cdot \vec{\sigma}$, with $\vec{k} = k_x \vec{e}_x + k_y \vec{e}_y$, assuming the rotation of the Pauli matrices $\sigma_y \to - \sigma_x$, $\sigma_x \to \sigma_y$. This convention facilitates a determination of the one-dimensional projections $H_R^{(jj')} = \alpha_R k \, \vec{e}\,^{(jj')} \cdot \vec{\sigma}$ of $H_R$ onto the triangle's sides by means of identifying $\vec{e}\,^{(21)} = \vec{e}_x$, $\vec{e}\,^{(32)} = \mathcal{R} (\beta - \pi) \vec{e}_x$, and $\vec{e}\,^{(13)} = \mathcal{R} (\pi - \beta) \vec{e}_x$, where $\beta = \arccos\left(\frac{L}{2L'}\right)$ and $\mathcal{R}$ is the rotation matrix with the corresponding angle argument. The equivalent representation $H_R^{(jj')} = \alpha_R k \, U\,^{(jj')} \sigma_x U\,^{(jj') \, \dagger}$, with $U^{(21)}=1$ and $U^{(32)} = U^{(13) \, \dagger} = e^{\frac{i}{2} (\pi - \beta) \sigma_z} \equiv U$ proves to be especially useful in the derivation of the effective model.}, the spin-orbit vector in the effective three-channel model has different angles to the the propagation directions $\vec{e}\,^{(jj')}$ along the segments from the contact $j'$ to the contact $j$ ($j,j'= 1,2,3$). For simplicity, we will consider an isosceles triangle with a horizontal side of the length $L$ and two oblique sides of the length $L'$. We  assume that both the $s$-wave superconductors and the semiconductors have spatially uniform parameters. In particular, the energy gap parameter $\Delta_0$ has the same value in all superconducting leads $j$, and their phases $\varphi_j$ can be tuned by external fluxes.

In the next step, we integrate out the extended components of the effective wire model described above (and sketched in the left panel of Fig. \ref{fig: model_sketch}) mapping it onto an effective three-site low-energy model (see the right panel of Fig. \ref{fig: model_sketch}), where sites represent the zero-dimensional contact points. Referring interested readers to details of our microscopic derivation in the Supplemental Material (SM) \cite{SM}, we state  the following effective $12\times 12$ Nambu Hamiltonian, that describes the low-lying excitations in the discussed device:
\begin{widetext}
    \begin{align}
        h_{\text{eff}}=\left( \begin{array}{ccc}
      \epsilon\tau_z+\Delta_0 e^{\frac{i}{2} \varphi_{1} \tau_z}\tau_x e^{-\frac{i}{2} \varphi_{1} \tau_z}  & -t\tau_z  e^{i \varphi_{SO} \sigma_x}  & -t'\tau_z   U^{\dagger} e^{-i \varphi_{SO}' \sigma_x} U \\
        -t\tau_z e^{-i \varphi_{SO} \sigma_x} & \epsilon\tau_z+\Delta_0 e^{\frac{i}{2} \varphi_{2} \tau_z}\tau_xe^{-\frac{i}{2} \varphi_{2} \tau_z}  &  -t'\tau_z  U  e^{i \varphi_{SO}' \sigma_x} U^{\dagger} \\
         -t'\tau_z U^{\dagger}  e^{i \varphi_{SO}' \sigma_x} U & -t'\tau_z   U  e^{-i \varphi_{SO}' \sigma_x} U^{\dagger} & \epsilon'\tau_z+\Delta_0 e^{\frac{i}{2} \varphi_{3} \tau_z}\tau_xe^{-\frac{i}{2} \varphi_{3} \tau_z}
    \end{array} \right). \label{H_eff_main}
    \end{align}
\end{widetext}
The model's on-site energy $\epsilon,\ \epsilon'$ and  hopping $t,\ t'$ parameters  are obtained from the Fermi-level projections of the full system's Green's function, which is evaluated at corresponding spatial points. The unitary matrix $U=i\sigma_{z}e^{-\frac{i}{2}\beta \sigma_{z}}$ takes into account the above-described direction's change of the quasiparticle propagation at each contact. The spin-orbit phases $\varphi_{SO}=k_R L$ and $\varphi_{SO}'=k_R L'$, expressed in terms of the Rashba momentum $k_R = m_W \alpha_R$, are accumulated during the propagation through the legs of the lengths $L$ and $L'$.

The effective Hamiltonian (\ref{H_eff_main}) possesses various symmetries, which explain features of its spectrum to be discussed in the following. First of all, $h_{\text{eff}} (\{ \varphi_j \})$ possesses the particle-hole symmetry $C=\tau_y \sigma_y K$, with $K$ denoting complex conjugation. It relates the Hamiltonian to itself at each fixed set of the phases $\{ \varphi_j\}$:
\begin{align}
    C h_{\text{eff}} (\{ \varphi_j \}) C^{-1} = - h_{\text{eff}} (\{ \varphi_j \} ).
    \label{particle-hole}
\end{align}

The time-reversal operator $T = i \sigma_y K$, such that $T^2= - 1$, relates $h_{\text{eff}}$ evaluated at the phases $\{ \varphi_j \}$ to its time-reversal partner at $\{- \varphi_j \}$:
\begin{align}
    T h_{\text{eff}} (\{ \varphi_j \}) T^{-1} =  h_{\text{eff}} (\{ -\varphi_j \} ).
    \label{time-reversal}
\end{align}
This symmetry implies the twofold Kramers degeneracy of a spectrum at the \emph{time-reversal invariant phases} $\varphi_j = 0 , \pm \pi$.

Moreover, the Hamiltonian (\ref{H_eff_main}) conserves the pseudo-spin operator
\begin{align}
\Sigma=&  \mathcal{U}  \frac{U_{SO} - \frac12  \text{tr} [U_{SO}] }{2 i \sqrt{1- \frac14 (\text{tr} [U_{SO}])^2}}  \mathcal{U}^{\dagger} = \Sigma^{\dagger},\quad \Sigma^2 = \frac14 ,
\label{Sigma_def}
\end{align}
where $U_{SO}=e^{i \varphi_{SO} \sigma_x} (U e^{i \varphi_{SO}' \sigma_x}U^{\dagger} ) (U^{\dagger} e^{i \varphi_{SO}' \sigma_x}  U)$ is the total non-abelian closed-loop spin-orbit phase accumulated along the triangle's circumference, and $\mathcal{U}=\text{diag}\{1,\ e^{-i \varphi_{SO} \sigma_x}U_{SO},\ U^{\dagger}e^{i \varphi_{SO}' \sigma_x}U\}$ is a unitary transformation acting trivially in the particle-hole space: $[\Sigma, \tau_{x, y, z, 0}]=0$. We refer to the SM \cite{SM} for the explicit demonstration of the commutation $[\Sigma , h_{\text{eff}} (\{ \varphi_j\})]=0$. It is important to note that the pseudo-spin conservation generally holds for arbitrary triangular geometry, not just for the presently discussed isosceles case, with a general expression for $\Sigma$ being more involved than in Eq. (\ref{Sigma_def}).

The pseudo-spin conservation implies that each ABS is characterized by the pseudo-spin quantum number ($+\frac12$ or $-\frac12$), and two ABS with opposite pseudo-spin values have a protected crossing. Moreover, a hole-like ABS has an opposite pseudo-spin value of its particle-like ABS partner --- this follows from the anticommutation 
$\{ C , \Sigma \} =0$. In the next Section \ref{transitions_main}, we show that pseudo-spin conservation prevents direct transitions between the pseudo-spin-split Andreev levels (current operators $\propto\tau_{x, y}$ commute with $\Sigma$). We show that to bypass these selection rules and couple two pseudo-spin states, a pseudo-spin-flipping element is required, similar to the two-terminal case studied in a recent theoretical work~\cite{Meyer2024}. One can achieve that by applying an external magnetic field or introducing a magnetic impurity.

\begin{figure}[h!]
\includegraphics[width=0.47\textwidth]{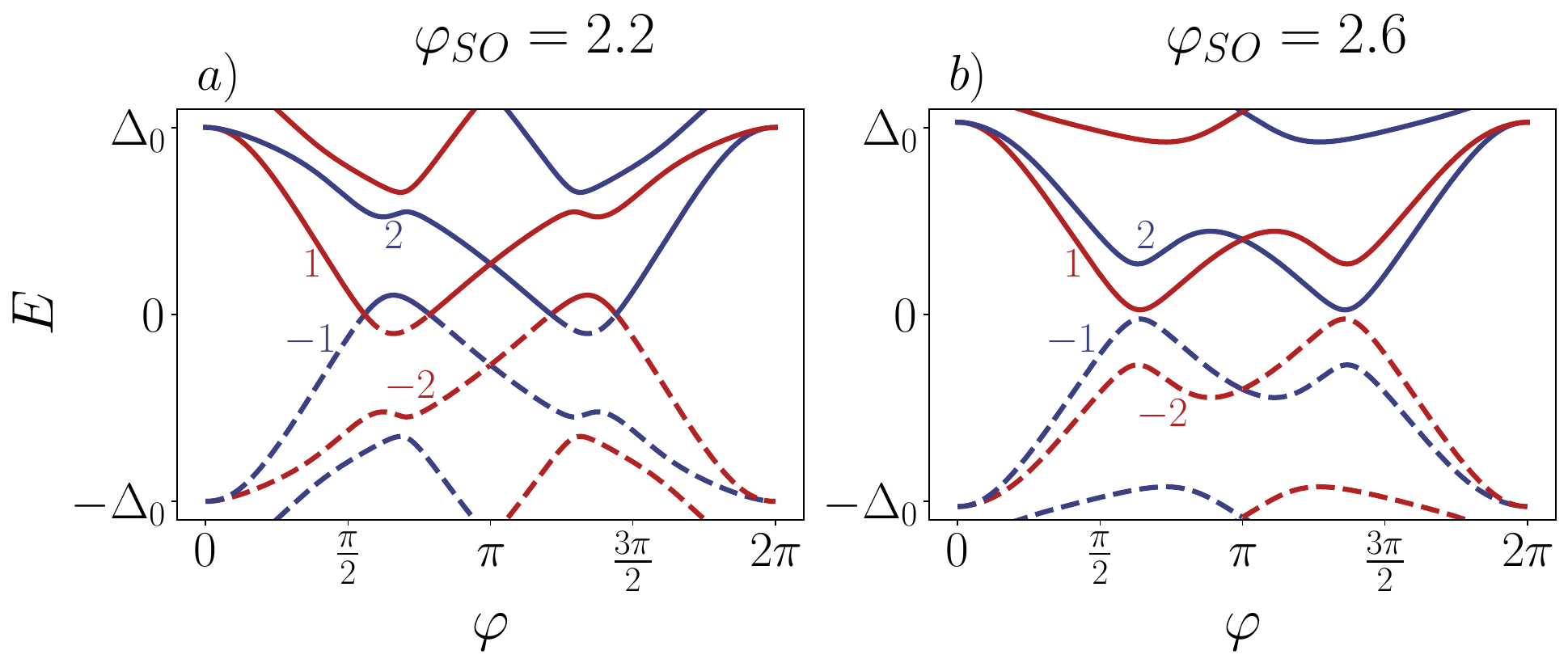}
\caption{The example of the energy spectrum of the effective Hamiltonian of Eq. \eqref{H_eff_main} for the case of an isosceles triangle $L=0.3L'$ with $\varphi_{1}=-\varphi_{2}=\varphi,\ \varphi_{3}=0$, $t=10t'/3=0.7\Delta_0,\ \epsilon=10\epsilon'/3=0.35\Delta_0$, and two distinct values of the spin-orbit angle $\varphi_{SO}=0.3\varphi_{SO}'=2.2,\ 2.6$. Red and blue lines represent two different pseudo-spin projections, while solid and dashed lines distinguish particle-like and hole-like solutions. Note, that the effective Hamiltonian provides quantitatively accurate results only in the vicinity of zero energy.}
               \label{fig: spectrum_example}
\end{figure}

Models belonging to this symmetry class (three superconducting contacts attached to a semiconducting island with spin-orbit coupling) may host zero-energy eigenstates. In Ref.~\cite{Akhmerov2014} the necessary condition for the existence of such eigenstates has been derived in the framework of the scattering matrix approach. It states that the interior of a triangle formed by the points $e^{i \varphi_1}$, $e^{i \varphi_2}$, $e^{i \varphi_3}$, which lie on the unit circle in the complex plane, must contain the circle's center. In the SM \cite{SM} we derive this condition for a spectrum of the effective Hamiltonian (\ref{H_eff_main}) by analyzing possibilities to satisfy the equation $\det \, [h_{\text{eff}} (\{\varphi_j\} )] =0$.

Applying the necessary condition for the choice of the phases $\varphi_3 =0$, $\varphi_1 = - \varphi_2 \equiv \varphi$, we deduce that zero-energy states may only exist at $\varphi \in [\frac{\pi}{2}, \frac{3 \pi}{2}]$. Plotting the spectrum of $h_{\text{eff}} (\{ \varphi, - \varphi ,0 \}) \equiv h_{\text{eff}} (\varphi)$ versus $\varphi$ in Fig.~\ref{fig: spectrum_example}, we observe that the zero-energy states do occur at the spin-orbit phase value $\varphi_{SO}=2.2$. Since the spectrum is particle-hole symmetric, a pair of two crossings at zero energy occurs on the interval of $\varphi$ between $\frac{\pi}{2}$ and $\pi$. The two crossings coalesce and disappear, as the parameter $\varphi_{SO}$ is continuously cranked up to the value $2.6$.

In addition to the zero-energy states at $\varphi < \pi$ we also have their symmetric partners at $\varphi > \pi$. This happens due to the above-made special choice of phases, which leads to an emergent symmetry of the spectrum concerning its reflection about the $\varphi = \pi$ point. Formally, this symmetry is expressed in terms of the unitary operator $\Pi=\sigma_{y}R$, with $\Pi^2 =1$,
\begin{align}
    \Pi h_{\text{eff}}(\varphi - \pi)\Pi^{-1}=h_{\text{eff}}(\pi-\varphi),
\end{align}
where the matrix
\begin{align}
    R=\begin{pmatrix}0 &1 & 0 \\ 1 & 0 & 0 \\ 0 & 0 & 1  \end{pmatrix}
\end{align}
geometrically reshuffles the contacts 1 and 2. It is also remarkable to note the emergence of two more symmetries supporting the spectral properties at a fixed value of $\varphi$: a) the anti-unitary time-reversal-like symmetry $\tilde{T}=-i \Pi T=RK$, with $\tilde{T}^2 =1$,
\begin{align}
    \tilde{T}  h_{\text{eff}} (\varphi) \tilde{T}^{-1} =  h_{\text{eff}} (\varphi),
\end{align}
and b) the unitary chiral symmetry $S=C \tilde{T} = \tau_y \sigma_y R$
\begin{align}
    S  h_{\text{eff}} (\varphi) S^{-1} = - h_{\text{eff}} (\varphi).
\end{align}

We find that the spectrum of our model is in good qualitative agreement with the one recently observed in the experiment \cite{Nichele2024}. In particular, the phase-dispersion of the lowest pair of spin-orbit split Andreev levels is described quite accurately, and we propose to utilize it for the realization of a TASQ. The values of $\varphi_{SO}$ used in the model are compatible with the sample sizes ($300\,\mathrm{nm}\times250\,\mathrm{nm}$) and estimated SOI length $k^{-1}_{R}\approx150\,\mathrm{nm}$.

To achieve an optimal regime for a single TASQ operation, the pair of positive-energy quasiparticle states should on one hand lie deep below $\Delta_0$, and on the other hand, it should feature a sufficiently large pseudo-spin splitting. In Fig.~\ref{fig: spectrum_example} we see that in the presence of zero energy crossings, a hole-like state turns into a particle-like state between the two successive crossings on the interval $0 < \varphi < \pi$ (and symmetrically on the interval $\pi < \varphi < 2 \pi$) retaining its pseudo-spin value. Thus, the two lowest positive-energy quasiparticle states between the crossings have the same pseudo-spin value, which is unfavorable for the realization of the ASQ. Ramping up the spin-orbit phase $\varphi_{SO}$, see Fig.~\ref{fig: spectrum_example}, we achieve the coalescence of the two crossings and their subsequent disappearance. After that, the two lowest positive energy states near $\varphi \approx \frac{2 \pi}{3}$ feature the opposite values of pseudo-spin as well as a sufficiently large pseudo-spin splitting. In addition, both phase dispersions (blue and red) feature local minima near this phase value, which protects the corresponding transition frequency against noise. This regime appears to be optimal for a single TASQ operation. Note that for two-qubit operations one needs to detune from that optimal point to have finite supercurrent (as local minima of phase dispersion correspond to zero supercurrent).

\section{Driving transitions and coupling two qubits}
\label{transitions_main}
The qubit can be operated by applying a current (flux)~\cite{Yeyati2017,Devoret2020,Devoret2021,Houzet2024} or a gate voltage drive~\cite{Nazarov2010,Tosi2019,Metzger2021,Yeyati2022,Heck2023}. Specifically, let us assume that the time-dependent voltage $V_{j}(t)$ is applied to the $j^{\text{th}}$ superconducting electrode. This, in turn, induces a temporal variation of the corresponding phase
\begin{align}
    \varphi_{j}\rightarrow \varphi_{j}(t)=\varphi_{j}+2\pi \frac{\Phi_{j}(t)}{\Phi_{0}},\quad \Phi_{j}(t)=\int^{t}dt'V_{j}(t'),
\end{align}
where $\Phi_{0}=\frac{\pi}{e},\ (\hbar=1)$ is the superconducting flux quantum. 

To the lowest order in the perturbation, the coupling to the external drive is thus described by the following Nambu Hamiltonian
\begin{align}
    H_{j}(t)=J_{j}\Phi_{j}(t), \label{drive_coupl}
\end{align}
where $J_{j}$ is the effective operator of the current flowing into the $j^{\text{th}}$ superconducting electrode:
\begin{align}
    J_{j}=&\frac{2\pi}{\Phi_{0}}\frac{\partial h_{\text{eff}}}{\partial \varphi_{j}}=-2\pi\frac{\Delta_0}{\Phi_{0}} e^{\frac{i}{2} \varphi_{j} \tau_z}\tau_{y} e^{-\frac{i}{2} \varphi_{j} \tau_z}\ket{j}\bra{j}.
\end{align}

\begin{figure}[t!]
\centering
      \includegraphics[width=0.45\textwidth]{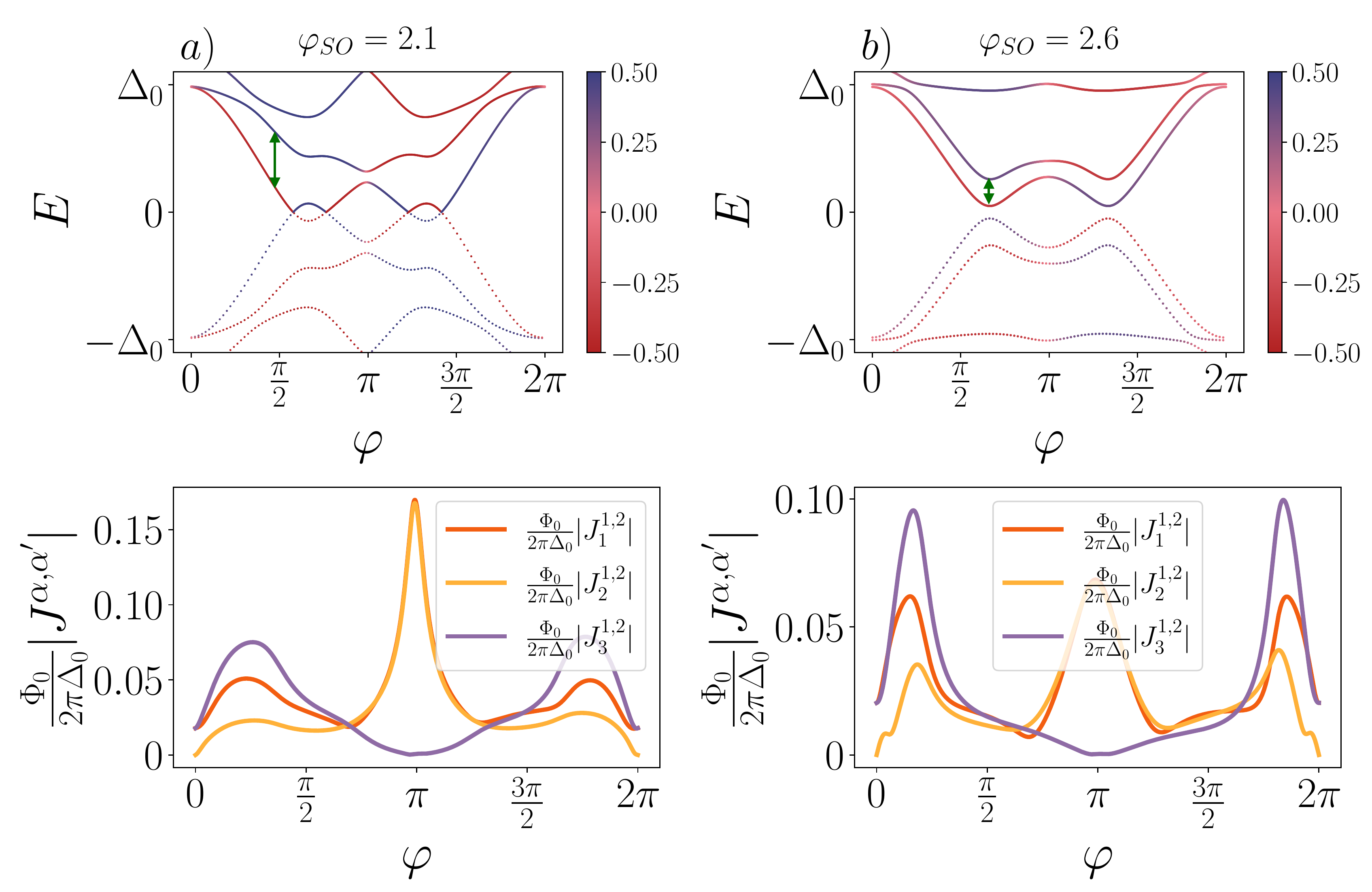}
                \caption{Numerical data for an equilateral junction ($L=L'$) in the in-plane magnetic field $\vec{B}=B(\cos\vartheta\vec{e}_{x}+\sin\vartheta\vec{e}_{y})$ with $\epsilon=0.32\Delta_0,\ t=0.7\Delta_0,\ \frac{g \mu_B  LB}{2 v_{F,S} \sin^2 (k_F L)}=0.1$ and $\vartheta=-\pi/3$. Panel a) shows the energy phase relation of the junction for two values of spin-orbit angle $\varphi_{SO}=2.1$ and $\varphi_{SO}=2.6$. The color map interpolates between the two distinct pseudo-spin projections: down-red and up-purple, while dashed and solid lines are used to differentiate between hole and particle-like bands. Panel b) shows the phase-dispersion of the current matrix elements between the lowest pair of pseudo-spin-split Andreev bound states, previously annulled by the pseudo-spin conservation.}
               \label{fig:trnasitions_mag}
\end{figure}

To study transitions between the relevant low-lying states $\alpha,\alpha' \in\{-2, -1, 1, 2\}$ as indicated in Fig. \ref{fig: spectrum_example}, we analyze the following matrix elements \footnote{Note that these matrix elements appear in the expansion of the second-quantized current operator $\widehat{J}_{j}$ into the Bogoliubov's quasiparticle operators $\gamma_{\alpha}, \gamma_{\alpha}^{\dagger}=\gamma_{-\alpha}$. In particular, the expression $\widehat{J}_{j}=\frac12 \sum_{\alpha, \alpha'}J_{j}^{\alpha, \alpha'}\gamma_{\alpha}^{\dagger}\gamma_{\alpha'}$ shall be used to get the correct interpretation of the allowed transitions in the system.}
\begin{align}
    J_{j}^{\alpha, \alpha'}=\braket{\alpha|J_{j}|\alpha'}
    \label{matrix_elem_nambu}
\end{align}
between the single particle states $h_{\text{eff}}| \alpha \rangle=E_{\alpha}| \alpha \rangle$ of the effective Hamiltonian in Eq. \eqref{H_eff_main}. In the first place, we notice that $[\Sigma , J_j] =0$, which implies that non-diagonal matrix elements between the states characterized by different pseudo-spin values vanish identically. The diagonal matrix elements, which can be expressed by the phase derivatives of the eigenenergies $J_{j}^{\alpha, \alpha}= \frac{2\pi}{\Phi_{0}}\frac{\partial E_{\alpha}}{\partial \varphi_j}$, provide an additional contribution to the supercurrent in the odd-parity states with the corresponding quasi-particle excitations.

To utilize the quasiparticle states $| \alpha \rangle$, with $\alpha =1,2$, which define the operational basis of the ASQ, one has to engineer a transition between the pseudo-spin eigenstates $1$ and $2$. For this purpose, it is necessary to violate the pseudo-spin conservation, which can be experimentally achieved, for example, by applying external magnetic fields. The former induces an effective Zeeman perturbation ${b}_{\text{eff}}$ (see SM~\cite{SM} for the specific form of this operator) to $h_{\text{eff}}$, which, in turn, does not commute with the pseudo-spin $[{b}_{\text{eff}}, \Sigma] \neq 0$, generating the required matrix element. Figure \ref{fig:trnasitions_mag} demonstrates this effect using a sample positioned in a weak in-plane magnetic field. Alternatively, in the absence of external magnetic fields, higher-energy levels can be used as a source of intermediate states for the required intra-doublet transition~\cite{Devoret2021,Yeyati2021}.

The initialization of the TASQ depends on a concrete realization. If the hybridized bound states are predominantly localized within the semiconductor, the odd state is easily achieved with gate voltage control. The preparation is more involved in the case of realization based on more delocalized ABSs. In that case, one can prepare the system in the odd-parity sector by a resonant microwave drive breaking a Cooper pair and ionizing one quasiparticle to the continuum~\cite{Heck2021,Kurilovich2024,zatsarynna2024many}.

\begin{figure}[t!]
                \includegraphics[width=0.4\textwidth]{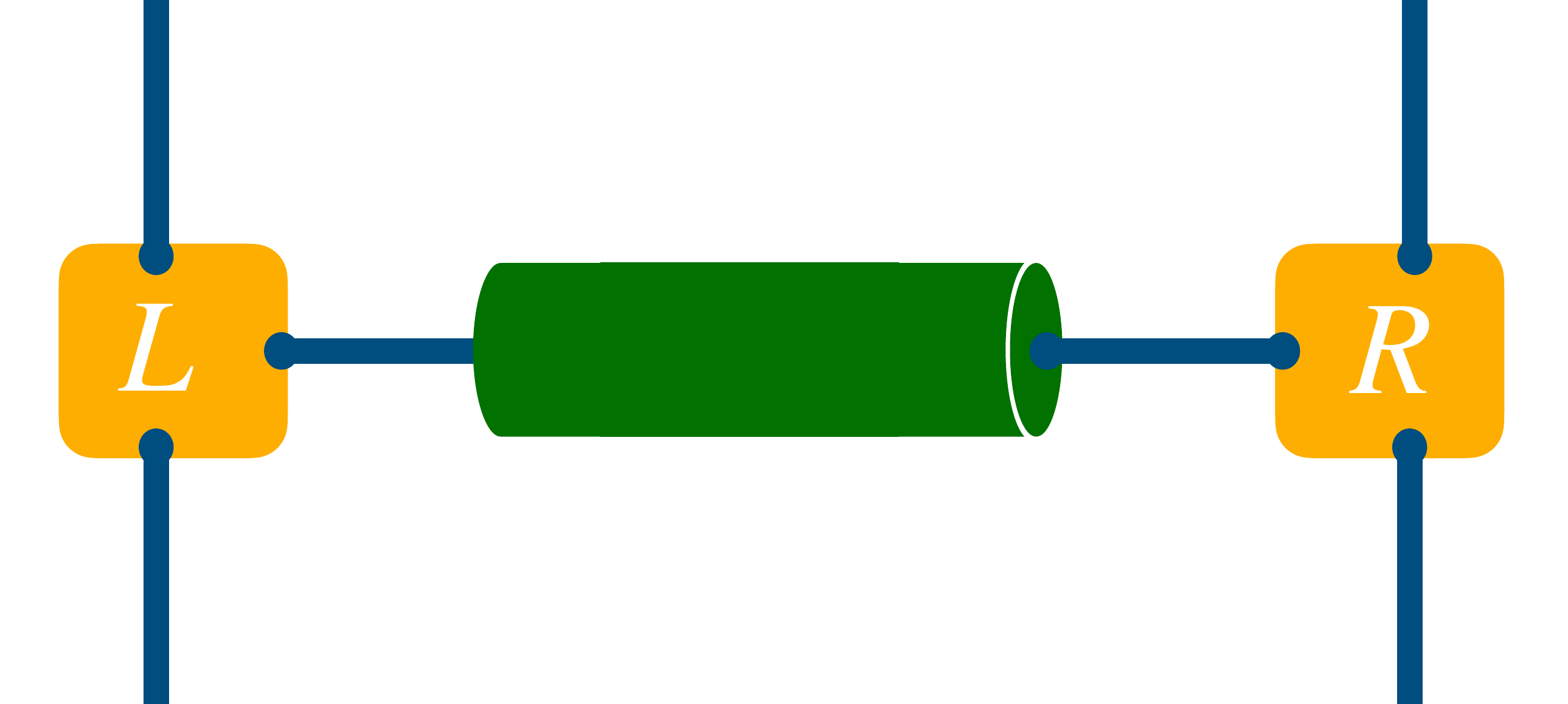}
                \caption{A schematic representation of the coupling between two TASQs mediated by the bosonic fluctuation of the shared phase variable inside the transmission line (shown in green).}
               \label{fig:Coupling}
\end{figure}
One can couple two tripod junctions via a superconducting circuit. Owing to the gapped nature of excitations in superconducting wires connecting the qubits, the direct overlap of ABS wave functions is exponentially suppressed. The leading order interaction between the qubits is provided by the fluctuations of the phase fields shared by the junctions~\cite{Nazarov2010,Andersen2024} (often referred to as inductive coupling).

In Fig. \ref{fig:Coupling} we present a sketch of the coupling between two TASQs mediated by bosonic fluctuation of the shared phase variable inside the transmission line. These fluctuations correspond to the propagating Mooij-Sch\"on plasma waves, which exhibit a linear, sound-like dispersion over a wide frequency range (as described in Ref. \cite{mooij1985propagating}). By integrating out the bosonic degrees of freedom (for technical details, see SM~\cite{SM}), we obtain the following effective two-qubit Hamiltonian, which captures the interaction between the qubits:
\begin{widetext}
    \begin{align}
    H_{\text{eff}}=\frac{1}{2}\sum_{S=L,R}\mathcal{E}_{S}\sigma_{z}^{(S)}-\mathcal{J}_{z}\sigma_{z}^{(R)}\sigma_{z}^{(L)}-\left(\mathcal{J}_{1}\sigma_{+}^{(R)}\sigma_{+}^{(L)}+\mathcal{J}_{1}^{*}\sigma_{-}^{(R)}\sigma_{-}^{(L)}+\mathcal{J}_{0}\sigma_{+}^{(R)}\sigma_{-}^{(L)}+\mathcal{J}_{0}^{*}\sigma_{+}^{(L)}\sigma_{-}^{(R)}\right)+\tilde{E}^{(0)} . \label{effectie_2qubit}
\end{align}
\end{widetext}
Here the Pauli operators $\sigma_{+}^{(S)}=\ket{e, S}\bra{g, S},\ \sigma_{-}^{(S)}=\ket{g, S}\bra{e, S}=\left(\sigma_{+}^{(S)}\right)^{\dagger}$ flip the quasiparticle states $\ket{g, S}=\hat{\gamma}_{1, S}^{\dagger}\ket{\text{GS}, S}$, $ \ket{e, S}=\hat{\gamma}_{2, S}^{\dagger}\ket{\text{GS}, S},$
defined for each qubit $S=L, R$. The qubits' transition frequencies  $\mathcal{E}_{S}$ (set by the energy difference $E_{2}^{(S)}-E_{1}^{(S)}$ in the non-interacting scenario), as well as the ground state energy $\tilde{E}^{(0)}$ (set by the (odd-parity) condensate energy $E_{1}^{(S)}-\frac{1}{2}\sum_{\alpha>0}E_{\alpha}^{(S)}$ in the non-interacting system) acquire interaction-induced corrections set by the total inductance $\ell$ of the transmission line connecting the subsystems (see the SM \cite{SM} for details). The exchange couplings $\mathcal{J}_{z},\ \mathcal{J}_{0},\ \mathcal{J}_{1}$ are estimated as $\mathcal{J}_{z}\sim \ell (J_{3, R}^{2,2}-J_{3, R}^{1,1})(J_{3, L}^{2,2}-J_{3, L}^{1,1})$, $\mathcal{J}_{0}\sim \ell J_{3, R}^{2,1}J_{3, L}^{1,2}$, and $\mathcal{J}_{1}\sim \ell J_{3, R}^{2,1}J_{3, L}^{2,1}$, where $J_{j, S}^{\alpha,\alpha'}$ is a matrix element of a current operator though a terminal $j$ of the qubit $S=L,R$ between the single-particle states $\alpha$ and $\alpha'$. In other words, the exchange couplings appear in the conventional form of an inductive current-current interaction $\frac{\ell J_{R}J_{L}}{2}$, well-known from classical circuit physics.

Further coupling the subsystems to an external deterministic drive (see Eq. \eqref{drive_coupl}), brings about the single-qubit transitions (see SM \cite{SM}): 
\begin{align}
    \left(J_{j, S}^{1, 2}\sigma_{-}^{(S)}+J_{j, S}^{2, 1}\sigma_{+}^{(S)}+\frac{J_{j, S}^{2, 2}-J_{j, S}^{1, 1}}{2}\sigma_{z}^{(S)}\right)\Phi^{(S)}_{j}(t),
\end{align}
which when taken together with the interactions of the Hamiltonian \eqref{effectie_2qubit}, enables implementing generic 2-qubit gates, essential for universal quantum computation \cite{divincenzo1995two}.

\section{Conclusions and outlook}
We have demonstrated, with a simple microscopic model, the possibility of utilizing a three-lead Josephson junction with SOI as a new type of a qubit -- TASQ. Our findings explain recent experimental observations~\cite{Nichele2024} and develop further the ideas proposed in a theoretical work~\cite{Akhmerov2014}. We have studied the possibility of driving one qubit and coupling two qubits.

To develop a useful qubit, one needs to create a good model to describe decoherence, as well as ways to mitigate it. Moreover, a realistic setup should be described with more involved models, including effects such as Coulomb and exchange interactions, disorder, and more than one conducting channel between each pair of leads. These issues should be addressed in future works.

\section*{Acknowledgments}

This work has greatly benefited from the Authors' discussions with Alexander Zazunov, Marco Coraiola, and Alexander Shnirman, for which we would like to express our gratitude.  KP acknowledges funding from Deutsche Forschungsgemeinschaft (DFG; German Research Foundation) Project SH 81/7-1. AS and WB acknowledge support by the Deutsche Forschungsgemeinschaft (DFG; German Research Foundation) via SFB 1432 (Project \mbox{No.} 425217212).

\bibliography{citations.bib}

\end{document}


\title{Theory of three-terminal Andreev spin qubits: Supplemental material}

\author{Kiryl Piasotski}
\email[Email: ]{kiryl.piasotski@kit.edu}
\affiliation{Institut f\"ur Theorie der Kondensierten Materie, Karlsruher Institut f\"ur Technologie, 76131 Karlsruhe, Germany}
\affiliation{Institut f\"ur Quanten Materialien und Technologien, Karlsruher Institut f\"ur Technologie, 76021 Karlsruhe, Germany}

\author{Aleksandr Svetogorov}
\affiliation{Universit\"at Konstanz, 78457 Konstanz, Germany}

\author{Wolfgang Belzig}
\affiliation{Universit\"at Konstanz, 78457 Konstanz, Germany}

\author{Mikhail Pletyukhov}
\affiliation{Institut für Theorie der Statistischen Physik, 
RWTH Aachen, 52074 Aachen, Germany}

\maketitle

\section{Derivation of the effective Hamiltonian}

\subsection{Mapping the semiconducting plate onto the wire model}
\label{Spin-orbit-explained}

\begin{figure}[b!]
                \includegraphics[width=0.4\textwidth]{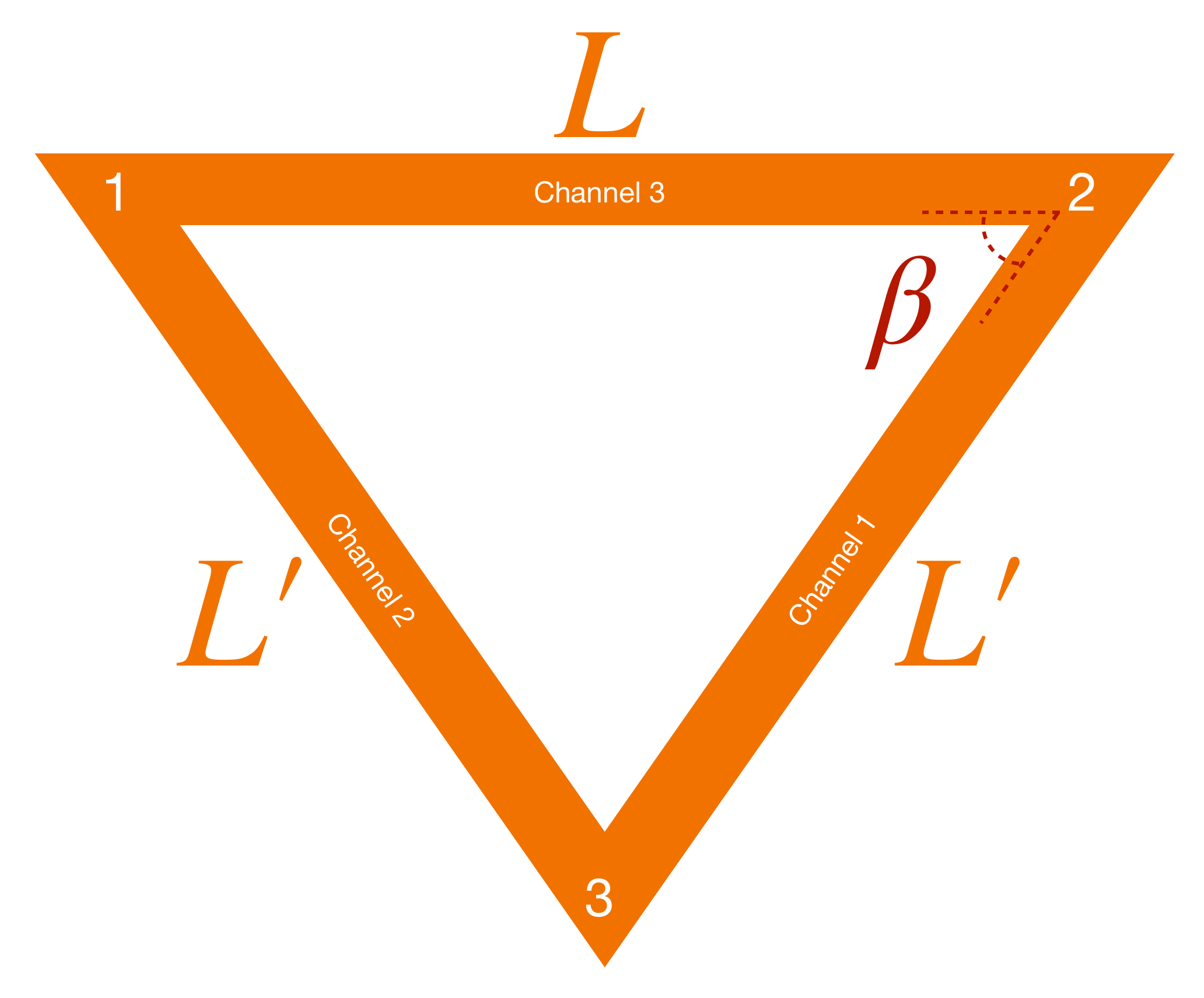}
                \caption{The sketch of the effective three-wire system.}
               \label{fig: 3_wire_sketch}
\end{figure}

Let us consider a two-dimensional (2D) electron gas described by the Hamiltonian (setting $\hbar =1$ in the following)
\begin{align}
    H=\frac{p_{x}^{2}+p_{y}^{2}}{2m_{W}}+\alpha_R \left(-\sigma_{y} p_{x}+\sigma_{x}p_{y}\right)-\mu_{W}\to \frac{p_{x}^{2}+p_{y}^{2}}{2m_{W}}+\alpha_R \left(\sigma_{x}p_{x}+\sigma_{y}p_{y}\right)-\mu_{W}, \label{2D_ham_semicond}
\end{align}
where $m_{W}, \mu_{W}$ are the effective mass and chemical potential, correspondingly. The spin-orbit velocity $\alpha_R $ gives rise to a new momentum (energy) scale $k_{R}=m_{W}\alpha_R$ ($E_{R}=\frac{m_{W}\alpha_R^{2}}{2}$), referred to as the spin-orbit Rashba momentum (energy) in the following. As explained in the main text, for the sake of convenience we perform a global unitary transformation rotating the Pauli matrices $\sigma_y \to - \sigma_x$, $\sigma_x \to \sigma_y$.

To model a propagation of quasiparticles between electrodes attached to the 2D semiconducting plate, we replace the latter by quasi-one-dimensional channels (wires) pairwise connecting the contacts. The wires form an isosceles triangle of side lengths $L$ and $L'$. To describe each of the wires, we project the Hamiltonian \eqref{2D_ham_semicond} onto an appropriate transverse mode perpendicular to the wire's direction. In particular, for the channel $3$ between the contacts $1$ and $2$ we choose the longitudinal axis along the $x$ axis, and the transverse one along the $y$ axis, as is shown in Fig. \ref{fig: 3_wire_sketch}. Averaging over the transverse mode gives $\langle p_y\rangle =0$, and thus we obtain the effective wire's Hamiltonian
\begin{align}
    H^{(21)}=\frac{(p_{x} + k_R \sigma_x)^{2}}{2m_{W}} + \frac{\langle p_y^2 \rangle}{2 m_W} - E_R -\mu_{W} = \frac{(p_{x} + k_R \sigma_x)^{2}}{2m_{W}} - \tilde{\mu}_W = e^{-i \sigma_x k_R x } \left( \frac{p_x^{2}}{2m_{W}} - \tilde{\mu}_W \right) e^{i \sigma_x k_R x } .
\end{align}

Assuming the open boundary conditions at the contacts $1$ ($x=0$) and $2$
 ($x=L$) we state the corresponding retarded Green's function of the finite-length wire
 \begin{align}
     G_W^{(21)} (x,x') = e^{-i \sigma_x k_R x} G_{W,0}^{(21)} (x,x') e^{i \sigma_x k_R x'},
     \label{GF_W}
 \end{align}
 where
\begin{align}
    G_{W,0}^{(21)} (x,x') = \frac{m_W}{k_0} \frac{\cos [k_0 (L- |x-x'|)] - \cos [k_0 (L-x-x')]}{\sin (k_0 L)} , \quad k_0 = \sqrt{2 m_W (\omega + i 0^+ + \tilde{\mu}_W)}
    \label{GW0}
\end{align}
is evaluated at $k_R=0$. Later on, we use the expression \eqref{GF_W} to reconstruct the coupling between the wires and the electrodes.

Likewise, we describe the two ballistic channels of length $L'$ connecting the electrodes $2\leftrightarrow3$ and $3\leftrightarrow1$. Their Hamiltonians are obtained by rotating the original coordinate system by the angles $\beta -\pi$ and $\pi - \beta$, respectively, with the acute angle $\beta = \arccos\left(\frac{L}{2L'}\right)$ as indicated in Fig. \ref{fig: 3_wire_sketch}. In terms of the local longitudinal coordinates $x_1$ and $x_2$ (chosen along the channels $1$ and $2$, respectively), they read
\begin{align}
    H^{(32)}&= U \frac{(p_{x_1} + k_R \sigma_x)^{2}}{2m_{W}} U^{\dagger} - \tilde{\mu}_W  , \label{H32} \\
    H^{(13)}&=U^{\dagger} \frac{(p_{x_2} + k_R \sigma_x)^{2}}{2m_{W}} U - \tilde{\mu}_W ,   \label{H13}
\end{align}
where  $U=e^{-\frac{i}{2} (\beta -\pi) \sigma_z}$. In the above derivation we used the relation
\begin{align}
    \vec{\sigma} \cdot \vec{p} = \sigma_x p_x + \sigma_y p_y = \sigma_{x'} p_{x'} + \sigma_{y'} p_{y'},
\end{align}
where $(x,y)$ and $(x',y')$ are the coordinates in the initial and the rotated local coordinate frames, respectively. Projecting the spin-orbit Hamiltonian onto the longitudinal direction $x'$ in the rotated frame (assuming $\langle p_{y'} \rangle =0$), we obtain
\begin{align}
    H'_R = \alpha_R \sigma_{x'} p_{x'} = \alpha_R (\vec{e}\,'_x \cdot \vec{\sigma})  p_{x'} = \alpha_R \left( \sum_{b=x,y,z} \sigma_b \mathcal{R}_{bx} (\theta) \right) p_{x'} = \alpha_R \left(\mathcal{R} (-\theta) \vec{\sigma} \right)_x p_{x'} ,
\end{align}
where $\mathcal{R} (\theta)$ is the rotation matrix by the angle $\theta$ about the $z$-axis, satisfying the orthogonality condition  $\mathcal{R}^T (\theta) = \mathcal{R}^{-1} (\theta) \equiv  \mathcal{R} (-\theta)$. Using the relation
\begin{align}
     e^{ -\frac{i}{2} \theta \sigma_z} \vec{\sigma}  e^{ \frac{i}{2} \theta \sigma_z} = \mathcal{R} (-\theta) \vec{\sigma},
\end{align}
we express
\begin{align}
     H'_R =e^{ -\frac{i}{2} \theta \sigma_z} \left( \alpha_R \sigma_x p_{x'}\right) e^{ \frac{i}{2} \theta \sigma_z} .
\end{align}
Choosing $\theta = \beta - \pi$ and $\theta = \pi - \beta$, we obtain \eqref{H32} and \eqref{H13}, respectively.

The retarded Green's functions corresponding to the Hamiltonians \eqref{H32} and \eqref{H13} read
\begin{align}
    G_W^{(32)} (x,x') &= U e^{-i \sigma_x k_R x} G_{W,0}^{(32)} (x,x') e^{i \sigma_x k_R x'} U^{\dagger} \label{GF_W_32}, \\
    G_W^{(13)} (x,x') &= U^{\dagger} e^{-i \sigma_x k_R x} G_{W,0}^{(13)} (x,x') e^{i \sigma_x k_R x'} U ,\label{GF_W_13} 
\end{align}
where $x,x'$ are understood from now on as the longitudinal coordinates in the corresponding local coordinate frames, and $G_{W,0}^{(32)} (x,x')= G_{W,0}^{(13)} (x,x')$ are obtained from \eqref{GW0} by replacing $L \to L'$.

For the subsequent application, we note the useful relations
\begin{align}
    -\frac{1}{8 m_W^2} \lim_{x \to 0^+} \partial_x^2 G_{W}^{(21)} (x,x) &= -\frac{1}{8 m_W^2} \lim_{x \to L^-} \partial_x^2 G_{W}^{(21)} (x,x) = -\frac{k_0}{2 m_W}  \cot k_0 L, \label{ddG00} \\
    \frac{1}{4 m_W^2} \lim_{x \to 0^+} \lim_{x' \to L^-} \partial_x \partial_{x'}  G_{W}^{(21)} (x,x') &= \frac{k_0}{2 m_W} \frac{e^{i \sigma_x k_R L}}{\sin k_0 L}, \label{ddG0L} \\
    \frac{1}{4 m_W^2} \lim_{x \to L^-} \lim_{x' \to 0^+} \partial_x \partial_{x'} G_{W}^{(21)} (x,x') &=  \frac{k_0}{2 m_W} \frac{e^{-i \sigma_x k_R L}}{\sin k_0 L}. \label{ddGL0}
\end{align}

In the following, we are going to attach superconducting leads to the vertices of the semiconducting triangle and to study sub-gap bound states at energies $\omega$ (relative to the Fermi level) satisfying the inequalities $|\omega| < \Delta_0 \ll \tilde{\mu}_W$, where $\Delta_0$ is the superconductor's energy gap. In the short junction limit $L \ll \xi $, where $\xi = v_{F,S}/\Delta_0$ is the superconductor's coherence length, we can neglect the $\omega$-dependence of the above-constructed normal-subsystem Green's functions approximating
\begin{align}
    k_0 \approx k_F \equiv \sqrt{2 m_W \tilde{\mu}_W}.
    \label{short_limit}
\end{align}

\subsection{Attaching superconducting leads}

In the next step, we describe a coupling of the triangle vertices --- one by one --- to three different superconducting leads which are modeled by quasi-one-dimensional semi-infinite wires. Each of them is characterized by the superconducting gap parameter $\Delta_0$ and by individual, contact-specific superconducting phase $\varphi_j$, $j=1,2,3$.

In Ref. \onlinecite{Piasotski2024} a general scheme of constructing the (retarded) Green's function of a layered system in terms of the Green's functions of its isolated constituents has been worked out. The developed approach is also applicable to the presently discussed geometry. In particular, the full-system Green's function $G (\vec{r}, \vec{r}\,')$ \footnote{We note that this function is effectively one-dimensional as it is defined on the six links of the planar graph --- three semiconducting segments forming the triangle and three superconducting segments which are attached one by one to each vertex of the triangle. Nevertheless, we use two-dimensional vectors to indicate the positions of the triangle's vertices to keep notations compact.} projected onto the contacts' coordinates $\vec{r}_j$ defines the matrix $\mathcal{G}$ in the contacts' space:
\begin{align}
    \mathcal{G}_{jj'} = G (\vec{r}_j, \vec{r}_{j'}).
\end{align}
Its inverse $d = \mathcal{G}^{-1}$ determines the correction $\delta \rho (\omega)$ to the global density of states due to the coupling between the subsystems (three semiconducting wires and three superconducting leads),
\begin{align}
    \delta \rho (\omega)= -\frac{1}{\pi}\text{Im}\frac{\partial}{\partial \omega}\ln \det\left[d(\omega)\right],
    \label{DOS}
\end{align}
which essentially depends on the superconducting phases $\varphi_j$. The matrix $d (\omega)$ also  allows one to determine the spectrum of bound states from the equation $\det [d (\omega)]=0$. The matrix elements $d_{jj'}$ are evaluated by the following expressions:
\begin{align}
    d_{11} &= - V_0 \tau_z  - \frac{\tau_z}{8 m_W^2} \lim_{x \to 0^+} \partial_x^2 G_W^{(21)} (x,x)- \frac{\tau_z}{8 m_W^2} \lim_{x \to (L')^-}  \partial_x^2 G_W^{(13)} (x,x) + i \frac{v_{F,S}}{2}  e^{\frac{i}{2} \varphi_1 \tau_z} F (z) e^{-\frac{i}{2} \varphi_1 \tau_z}, \\
    d_{22} &= - V_0 \tau_z - \frac{\tau_z}{8 m_W^2} \lim_{x \to 0^+} \partial_x^2 G_W^{(32)} (x,x)- \frac{\tau_z}{8 m_W^2} \lim_{x \to L^-}  \partial_x^2 G_W^{(21)} (x,x) + i  \frac{v_{F,S}}{2} e^{\frac{i}{2}  \varphi_2 \tau_z} F (z) e^{-\frac{i}{2}  \varphi_2 \tau_z}, \\
    d_{33} &= - V_0 \tau_z - \frac{\tau_z}{8 m_W^2} \lim_{x \to 0^+} \partial_x^2 G_W^{(13)} (x,x)- \frac{\tau_z}{8 m_W^2} \lim_{x \to (L')^-}  \partial_x^2 G_W^{(32)} (x,x)  + i \frac{v_{F,S}}{2} e^{\frac{i}{2} \varphi_3 \tau_z} F (z) e^{-\frac{i}{2}  \varphi_3 \tau_z}, \\
    d_{12} &= (d_{21})^{\dagger} = \frac{\tau_z}{4 m_W^2} \lim_{x \to 0^+} \lim_{x' \to L^-} \partial_x \partial_{x'} G_W^{(21)} (x,x'), \\
    d_{23} &= (d_{32})^{\dagger} =\frac{\tau_z}{4 m_W^2} \lim_{x \to 0^+} \lim_{x' \to (L')^-} \partial_x \partial_{x'} G_W^{(32)} (x,x'), \\
    d_{31} &=(d_{13})^{\dagger} = \frac{\tau_z}{4 m_W^2} \lim_{x \to 0^+} \lim_{x' \to (L')^-} \partial_x \partial_{x'} G_W^{(13)} (x,x'), 
\end{align}
where $V_0$ is the amplitude of the contact potentials $v_j (\vec{r}) = V_0 \delta (\vec{r} - \vec{r}_j)$ \footnote{Again, this delta-function is effectively one-dimensional, even though its argument is given by a two-dimensional vector. A rigorous specification of the argument's form requires the usage of six local coordinate systems for each link of the graph yielding a bulky expression.}. The matrix $\tau_z$ augments the normal-subsystem Hamiltonian to the extended Nambu representation (at $B=0$). An effect of the superconducting leads is accounted for by the function
\begin{align}
    F (z= \omega + i 0^+) = \frac{1- \frac{\Delta_0}{z} \tau_x}{\sqrt{1- \left( \frac{\Delta_0}{z} \right)^2}} \stackrel{|\omega| < \Delta_0}{=}  \frac{\omega + i 0^+ - \Delta_0 \tau_x}{i\sqrt{ \Delta_0^2 - \omega^2}}, \label{SC_self_energy}
\end{align}
which is dressed by the superconducting phases $\varphi_j$. The factor $v_{F,S}$ is the Fermi velocity of the superconductors.

\begin{figure}[h!]
                \includegraphics[width=0.8\textwidth]{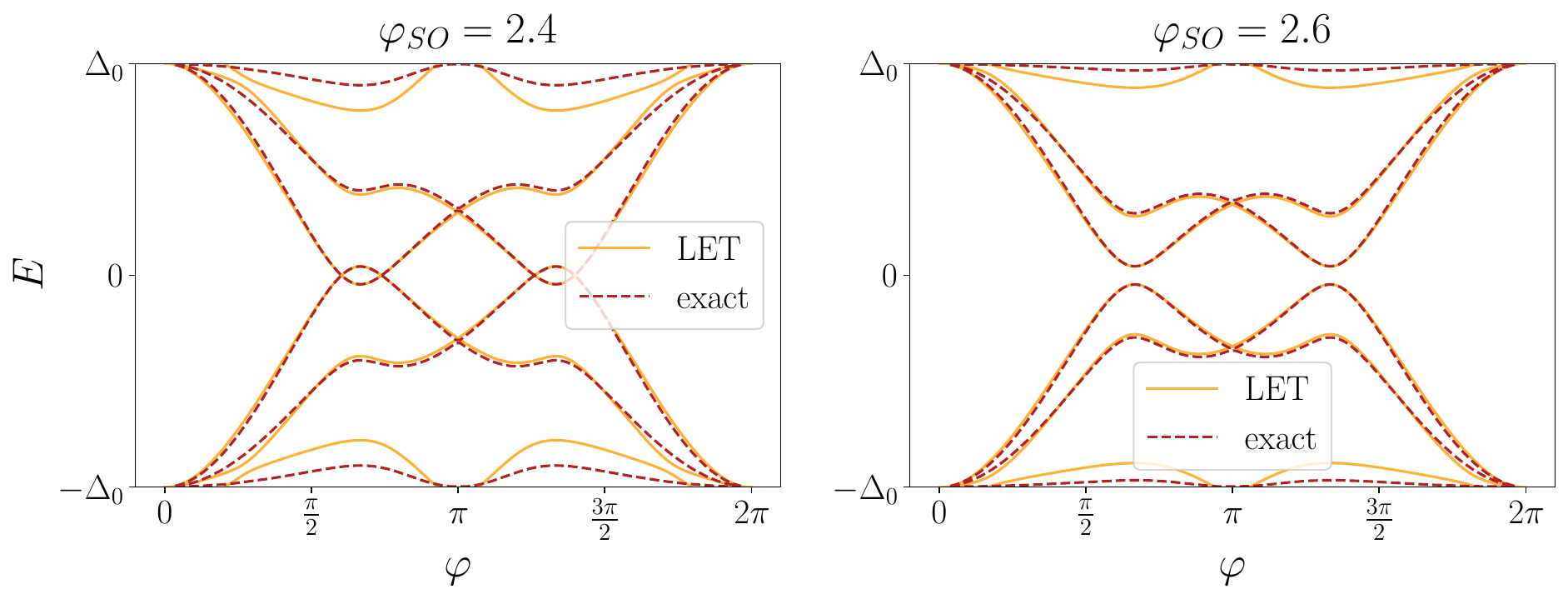}
                \caption{The comparison of the \textit{low-energy theory (LET)} spectrum of the effective Hamiltonian \eqref{eff_ham_sm} with the solutions of the bound-state equation $\det [d (\omega)]=0$ \textit{(labeled as "exact")} which  retains the energy dependence of the superconductors' contribution (encoded in the function \eqref{SC_self_energy}) to the system's Green's function. The model's parameters here are $\epsilon=\epsilon' =0.75\Delta_0,\ t=t'=0.95\Delta_0$, and $L=L'$. As one can see, the effective Hamiltonian description ("LET") captures quite accurately the "exact" energy values of the lowest pair of the spin-split Andreev bound states that we propose to realize in the TASQ. For the description of the higher-lying excitations, including the continuum of scattering states, one has to relax the LET approximation \eqref{LET_appr_d}. We further remark that beyond  the short-junction limit one is also obliged to take into account the frequency dependence of the Green's functions of the wires by relaxing the approximation \eqref{short_limit}. }
               \label{fig: compar}
\end{figure}

Approximating at $\omega \ll \Delta_0$
\begin{align}
    F (\omega + i 0^+) \approx - \frac{i}{\Delta_0} \left[ \omega + i 0^+ - \Delta_0 \tau_x \right],
\end{align}
we represent
\begin{align}
    d (\omega) \approx \frac{v_{F,S}}{2 \Delta_0} [\omega + i 0^+ - h_{\text{eff}}],
    \label{LET_appr_d}
\end{align}
where the effective low-energy Hamiltonian 
\begin{align}
    h_{\text{eff}} = \left(  \begin{array}{ccc}
       \epsilon \tau_z + \Delta_0  e^{\frac{i}{2}  \varphi_1 \tau_z} \tau_x e^{-\frac{i}{2}  \varphi_1 \tau_z}  & -t \tau_z e^{i \varphi_{SO} \sigma_x}  & -t' \tau_z U^{\dagger} e^{-i \varphi'_{SO} \sigma_x} U \\
         -t \tau_z e^{-i \varphi_{SO} \sigma_x} & \epsilon \tau_z + \Delta_0  e^{\frac{i}{2}  \varphi_2 \tau_z} \tau_x e^{-\frac{i}{2} \varphi_2 \tau_z } & -t' \tau_z U e^{i \varphi'_{SO} \sigma_x} U^{\dagger} \\
         -t' \tau_z U^{\dagger} e^{i \varphi'_{SO} \sigma_x} U & -t' \tau_z U e^{-i \varphi'_{SO} \sigma_x} U^{\dagger} & \epsilon' \tau_z + \Delta_0 e^{\frac{i}{2}  \varphi_3 \tau_z} \tau_x e^{-\frac{i}{2}  \varphi_3 \tau_z}
    \end{array}\right) \label{eff_ham_sm}
\end{align}
is expressed in terms of the parameters $\varphi_{SO}=k_R L$, $\varphi'_{SO} = k_R L'$, and
\begin{align}
    \epsilon &= \frac{2 \Delta_0}{v_{F,S}} \left[ V_0 + \frac{k_F}{2 m_W} \cot (k_F L) + \frac{k_F}{2 m_W} \cot (k_F L') \right], \\
    \epsilon' &= \frac{2 \Delta_0}{v_{F,S}} \left[ V_0 + \frac{k_F}{m_W} \cot (k_F L')\right], \\
    t &= \frac{2 \Delta_0}{v_{F,S}} \frac{k_F}{2 m_W} \frac{1}{\sin (k_F L)}, \\
    t' &= \frac{2 \Delta_0}{v_{F,S}} \frac{k_F}{2 m_W} \frac{1}{\sin (k_F L')}.
\end{align}
The latter parameters are evaluated with the help of \eqref{ddG00}-\eqref{ddGL0}. The low-energy spectrum is then naturally determined from the equation
\begin{align}
    \det [\omega - h_{\text{eff}}] =0.
\end{align}

\section{Conserved pseudo-spin operator $\Sigma$}
\label{psseudospin}
Considering the hopping part of the Hamiltonian $h_{\text{eff}}$
\begin{align}
        h_{T}=\left( \begin{array}{ccc}
      0  & -t\tau_z  e^{i \varphi_{SO} \sigma_x}  & -t'\tau_z  U^{\dagger}  e^{-i \varphi_{SO}' \sigma_x} U \\
        -t\tau_z e^{-i \varphi_{SO} \sigma_x} & 0  &  -t'\tau_z  U  e^{i \varphi_{SO}' \sigma_x} U^{\dagger} \\
         -t'\tau_z U^{\dagger}  e^{i \varphi_{SO}' \sigma_x} U & -t'\tau_z   U  e^{-i \varphi_{SO}' \sigma_x} U^{\dagger} & 0
    \end{array} \right),
\end{align}
we apply the transformation 
\begin{align}
    \tilde{h}_{T}=\mathcal{U}^{\dagger}h_{T}\mathcal{U}=\begin{pmatrix}  0  & -t\tau_z  U_{SO}  & -t'\tau_z   \\
        -t\tau_z U_{SO}^{\dagger} & 0  &  -t'\tau_z     \\
         -t'\tau_z  & -t'\tau_z     & 0  \end{pmatrix} ,\quad 
         U_{SO}=\underbrace{e^{i \varphi_{SO} \sigma_x}}_{\equiv A} (\underbrace{U e^{i \varphi_{SO}' \sigma_x}U^{\dagger}}_{\equiv B} ) (\underbrace{U^{\dagger} e^{i \varphi_{SO}' \sigma_x}  U}_{\equiv C}), \label{transformed}
\end{align}
which is given by
\begin{align}
    \mathcal{U}=\begin{pmatrix}1 & & \\ & B C & \\ & & C \end{pmatrix}.
\end{align}
Note that $\mathcal{U}$ commutes with the diagonal part of $h_{\text{eff}}$.

The Hamiltonian  \eqref{transformed} has a conserved pseudo-spin $\tilde{\Sigma}$, that is the generator $\tilde{\Sigma}=\frac12 \vec{n}_{\Sigma}\cdot \vec{\sigma}$ of the unitary transformation $U_{SO}=e^{i \frac{\theta_{\Sigma}}{2} \vec{n}_{\Sigma}\cdot \vec{\sigma}} = \cos \frac{\theta_{\Sigma}}{2} + 2 i \tilde{\Sigma} \sin \frac{\theta_{\Sigma}}{2}$.
It follows
\begin{align}
    \tilde{\Sigma} = \frac{U_{SO} - \frac12  \text{tr} [U_{SO}] }{2 i \sqrt{1- \frac14 (\text{tr} [U_{SO}])^2}}.
\end{align}
Hence the original Hamiltonian $h_{\text{eff}}$ conserves the pseudo-spin
\begin{align}
    \Sigma =  \mathcal{U}  \frac{U_{SO} - \frac12  \text{tr} [U_{SO}] }{2 i \sqrt{1- \frac14 (\text{tr} [U_{SO}])^2}}  \mathcal{U}^{\dagger} = \frac{\mathcal{U}_{SO} - \frac12  \text{tr} [U_{SO}] }{2 i \sqrt{1- \frac14 (\text{tr} [U_{SO}])^2}} , \quad \mathcal{U}_{SO} = \begin{pmatrix} A B C & & \\ & BC A & \\ & & CAB  \end{pmatrix} .
\end{align}
In particular, for the equilateral triangle with $L=L'$, $U^3 =-1$, and $U_{SO} = -\left( e^{i \varphi_{SO} \sigma_x} e^{i \frac{\pi}{3} \sigma_z} \right)^3$, this expression simplifies to
\begin{align}
    \Sigma=& \begin{pmatrix}\tilde{\Sigma} & & \\ & U\tilde{\Sigma} U^{\dagger} & \\ & & U^{\dagger} \tilde{\Sigma} U\end{pmatrix},\quad \tilde{\Sigma}=\frac{\sin\varphi_{SO}\frac{\sigma_{x}+\sqrt{3}\sigma_{y}}{2}+\cos\varphi_{SO} \frac{\sqrt{3}}{2} \sigma_{z}}{2 \sqrt{1-\frac{1}{4}\cos^{2}\varphi_{SO}}} .
\end{align}

It is remarkable that at vanishing spin-orbit interaction $\varphi_{SO} \to 0$, the pseudo-spin turns into the physical spin. Expanding
\begin{align}
    U_{SO} & \approx \left(1 + i \varphi_{SO} \sigma_x - \frac12 (\varphi_{SO})^2 \right) U \left(1 + i \varphi'_{SO} \sigma_x - \frac12 (\varphi'_{SO})^2 \right) (U^{\dagger})^2 \left(1 + i \varphi'_{SO} \sigma_x - \frac12 (\varphi'_{SO})^2 \right) U  \\
    & \approx 1 + i \varphi_{SO} \left[ \sigma_x +  \frac{L'}{L} U \sigma_x U^{\dagger}+  \frac{L'}{L} U^{\dagger} \sigma_x U \right] \nonumber \\
    &- \varphi_{SO}^2 \left[  \frac12 + \frac{(L')^2}{L^2}+ \frac{L'}{L} \sigma_x U \sigma_x U^{\dagger} + \frac{L'}{L} \sigma_x U^{\dagger} \sigma_x U + \frac{(L')^2}{L^2} U \sigma_x (U^{\dagger})^2 \sigma_x U \right] \\
    &= 1 + i \varphi_{SO} \sigma_x \left[ 1 - \frac{2 L'}{L} \cos \beta \right] - \varphi_{SO}^2 \left[  \frac12 + \frac{(L')^2}{L^2} - \frac{2 L'}{L} \cos \beta+ \frac{(L')^2}{L^2} e^{-2 i \beta \sigma_z} \right] \\
    & = 1 + \frac{i}{2} \varphi_{SO}^2   \tan \beta \, \sigma_z \approx e^{\frac{i}{2} \varphi_{SO}^2   \tan \beta \, \sigma_z},
\end{align}
we identify
\begin{align}
    \lim_{\varphi_{SO} \to 0 } \tilde{\Sigma} = \frac12 \sigma_z ,
\end{align}
as well as $\Sigma \to \frac12 \sigma_z$ in the limit of vanishing spin-orbit interaction.

\section{Necessary condition for the existence of zero-energy solutions}
\label{zero_energy_states}

To establish a necessary condition for zero-energy solutions, we analyze the equation $\det\left[h_{\text{eff}}\right]=0$. It can also be written as
\begin{align}
    \det\left[h_{\text{N}}+\Delta_{0} e^{i\check{\varphi}\tau_{z}/2}\tau_{x}e^{-i\check{\varphi}\tau_{z}/2}\right]=0,  \label{zero_energy_line1} 
\end{align}
where $h_{\text{N}}=\tau_{z}\check{h}_{\text{N}}$ is the normal part of the Hamiltonian obtained from $h_{\text{eff}}$ by setting $\Delta_{0}=0$ (and $B=0$), and
\begin{align}
    \check{\varphi}=\sum_{j=1}^3 \varphi_{j}\ket{j}\bra{j}.
\end{align}
The matrix under the determinant operation in Eq. \eqref{zero_energy_line1} can be reduced in size, effectively undoing the Nambu doubling of components. This leads to the reformulation of the same equation in the form
\begin{align}
\det\left[\check{h}_{\text{N}}e^{i\check{\varphi}}\check{h}_{\text{N}}+\Delta_{0}^{2} e^{i\check{\varphi}}\right]=0. \label{zero_en_eq}
\end{align}

First, we note that for all $\varphi_{j} =0$ it is impossible to satisfy Eq. \eqref{zero_en_eq}: The matrix $  \check{h}_{\text{N}}^2 + \Delta_{0}^{2}$ is positive definite, and therefore all its eigenvalues are strictly greater than zero, yielding $\det  \left[ \check{h}_{\text{N}}^2 + \Delta_{0}^{2}  \right] >0$.

Next, we introduce hermitian matrices
\begin{align}
    Q_{j} = \sum_{\sigma}\left( \check{h}_{\text{N}}  | j , \sigma \rangle \langle j , \sigma |   \check{h}_{\text{N}} + \Delta_{0}^{2}  | j , \sigma \rangle \langle j , \sigma | \right), 
\end{align}
in terms of the basis states $|j , \sigma \rangle$, which are labeled by the site index $j=1,2,3$ and the spin index $\sigma$. These matrices are positive semi-definite, since for any state $| \psi \rangle $ with a nonzero norm it holds
\begin{align}
    \langle \psi | Q_{j} | \psi \rangle  =  \sum_{\sigma}\left(  | \langle j , \sigma |   \check{h}_{\text{N}} | \psi \rangle |^2+ \Delta^{2}_{0}  | \langle j , \sigma  | \psi \rangle |^2\right) \geq 0 .
\end{align}

The matrix $\sum_{j} Q_{j} = \check{h}_{\text{N}}^2 + \Delta^{2}_{0}$ is positive definite though, as explained above.

\begin{figure}[h!]
\centering
                \includegraphics[width=0.8\textwidth]{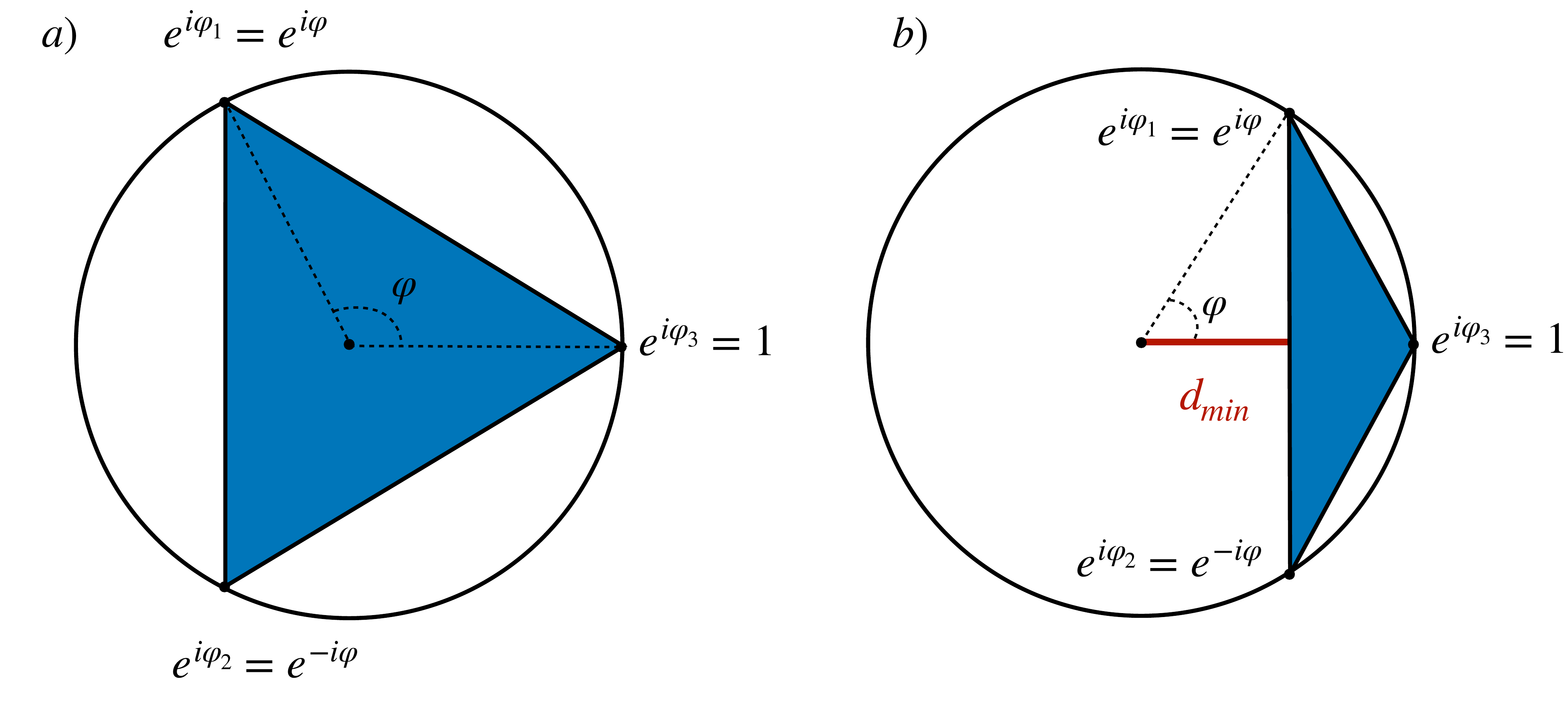}
                \caption{The points inside (and on the boundary) of the blue triangle are parametrized by affine combinations $w_{1}e^{i\varphi_{1}}+w_2 e^{i \varphi_2} +w_{3}e^{i\varphi_{3}}$ with non-negative weights $w_{j}$ satisfying the normalization  $\sum_{j}w_{j}=1$. Here, we demonstrate a specific choice of phases  $\varphi_{1}=-\varphi_{2}=\varphi,\ \varphi_{3}=0$ addressed in the main text. a) When the triangle encloses the circle's center, zero-energy solutions of the equation $\det [\omega - h_{\text{eff}}]=0$ may occur. An arrangement of phases with the enclosed circle's center represents the necessary (but not sufficient) condition for the existence of zero-energy solutions. b)  When the triangle does not enclose the circle's center, zero-energy solutions do not occur, since the condition $\det [ h_{\text{eff}}]=0$ can not be satisfied at any choice of the model's parameters.}
               \label{fig:polygon}
\end{figure}

Rewriting the equation \eqref{zero_en_eq} in terms of $Q_{j}$ yields
\begin{align}
    \det \left[ \sum_{j} Q_{j} e^{i \varphi_{j}}\right] =0.
    \label{detQj}
\end{align}
Let us study eigenvalues of the non-hermitian matrix $\sum_{j} Q_{j} e^{i \varphi_{j}}$ and try to establish a lower bound on their absolute values. We set up the eigenvalue equation
\begin{align}
     \sum_{j} Q_{j} e^{i \varphi_{j}} | \chi  \rangle = \lambda | \chi  \rangle ,
     \label{Qj_EV}
\end{align}
where $\lambda$ is in general complex-valued. Multiplying \eqref{Qj_EV} it with $\langle \chi  | = [ | \chi \rangle ]^{\dagger}$ (note that it is not a left eigenvector of $\sum_{j} Q_{j} e^{i \varphi_{j}}$, since this matrix is non-hermitian!), we obtain
\begin{align}
    \sum_{j} \tilde{w}_{j} e^{i \varphi_{j}}  = \lambda \langle \chi  | \chi  \rangle ,\quad 
    \tilde{w}_{j} = \langle \chi |Q_{j} | \chi\rangle  \geq 0 ,
\end{align}
with $\langle \chi | \chi \rangle = |\chi|^2 > 0$. In turn,
\begin{align}
    N= \sum_{j} \tilde{w}_{j} =  \langle \chi | \sum_{j} Q_{j} | \chi\rangle >0 .
\end{align}
The latter strict inequality allows us to define weights $w_j = \frac{\tilde{w}_{j}}{N} \geq 0$ such that $\sum_{j} w_{j} =1$, and thus we obtain
\begin{align}
     \big| \sum_{j} w_{j} e^{i \varphi_{j}} \big| = | \lambda | \frac{|\chi|^2}{N} . \label{key_eqn_zero_energy}
\end{align}

The convex set of points on the complex plane
\begin{align}
    P=\left\{\sum_{j} w_{j} e^{i \varphi_{j}}\,\,\, \Big{|} \quad w_{j}\geq0, \quad \sum_{j} w_{j}=1\right\}
\end{align}
forms a polygon's interior \cite{lang2012introduction}, with the polygon's vertices $\left\{e^{i \varphi_{j}}\right\}_{j}$ lying on the unit circle (that is, $P$ is the convex hull of the set $\left\{e^{i \varphi_{j}}\right\}_{j}$). In this connection, the left-hand side of Eq. \eqref{key_eqn_zero_energy} may only become zero, if the area of the polygon includes the circle's center, as illustrated in Fig. \ref{fig:polygon}(a), meaning that the distance from the circle's center to $P$ may be in principle annulled by some choice of $w_j$. In turn, if the circle's center lies outside of the polygon's area, like in Fig. \ref{fig:polygon}(b), then the shortest distance $d_{\text{min}}$ from the center to $P$ is the distance from the center to the nearest chord. This gives rise to the inequality
\begin{align}
    \big| \sum_{j} w_{j} e^{i \varphi_{j}} \big| \geq d_{\text{min}} ,
\end{align}
which implies the existence of the lower bound 
\begin{align}
    |\lambda | \geq \frac{N d_{\text{min}}}{|\chi|^2} >0 
\end{align}
for $|\lambda|$. Thus, neither eigenvalue of the matrix $\sum_j Q_j e^{i \varphi_j}$ is zero-valued, and the equations \eqref{detQj} and \eqref{zero_energy_line1} can not be satisfied for any choice of $w_j$.

Thus, we recover the necessary condition for the existence of zero-energy solutions, which was initially given in Ref. \onlinecite{Akhmerov2014}, that the circle's center must be included in the polygon's interior, providing its alternative proof in the Hamiltonian formulation. The proof is straightforwardly generalized to an arbitrary number $N_t$ of terminals in a multi-terminal Josephson junction in the absence of magnetic field and/or other time-reversal symmetry-breaking interactions.

\section{Effect of magnetic field}
\label{magnetic field}

\begin{figure}[b!]
                \subfigure[]{\includegraphics[width=0.45\textwidth]{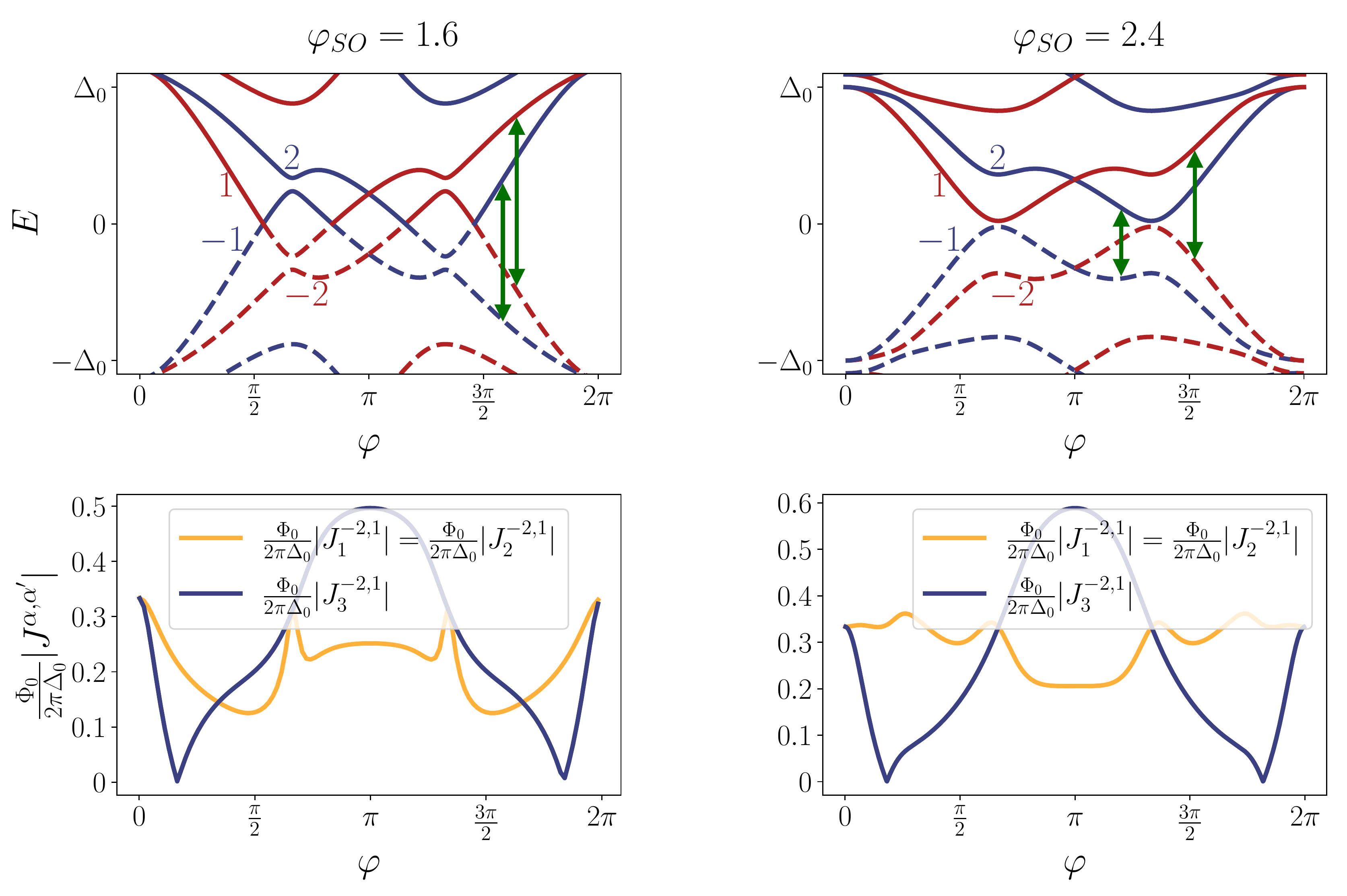}}
                \subfigure[]{\includegraphics[width=0.45\textwidth]{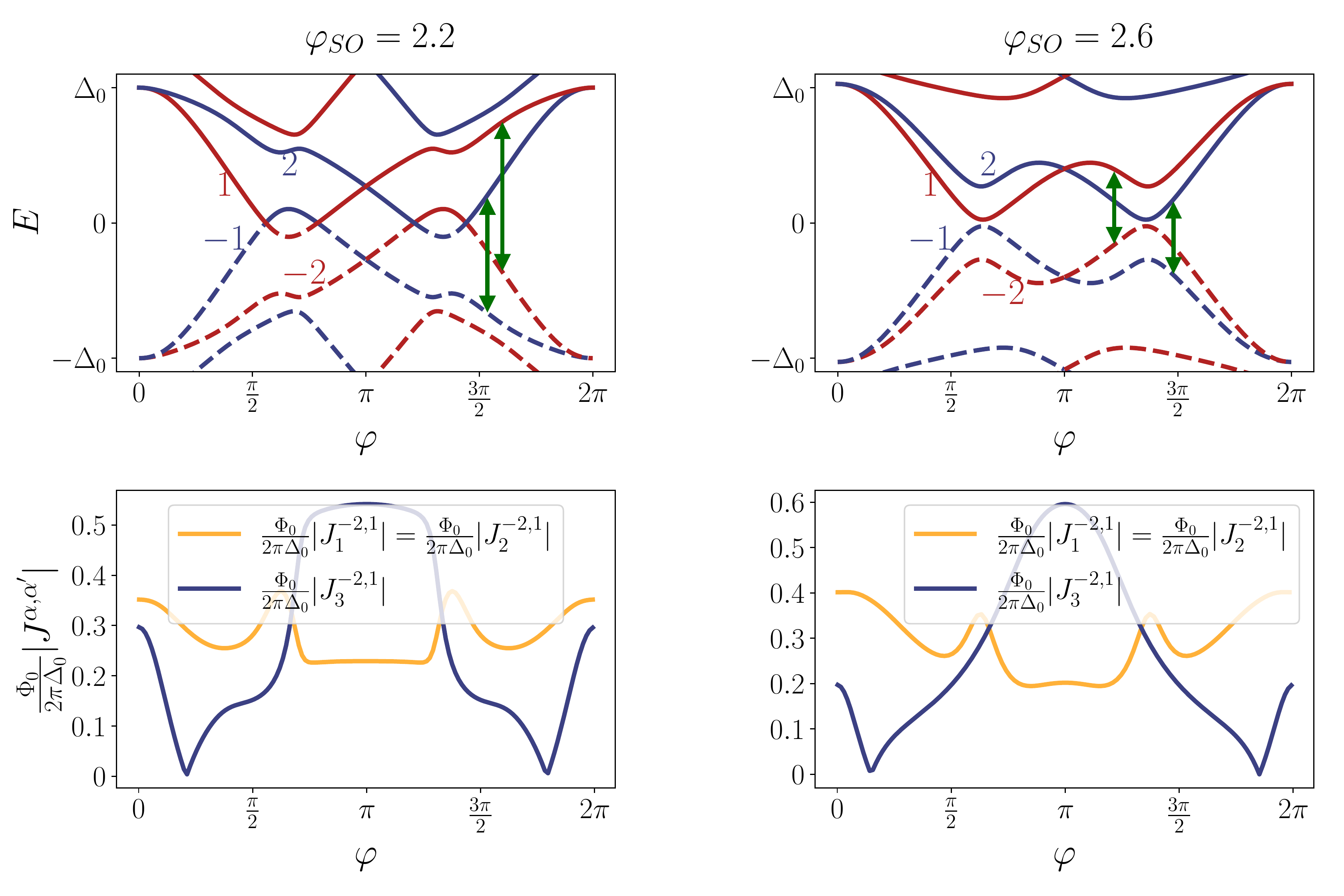}}
                \caption{The non-zero matrix elements between the low-lying states. Here sub-figures (a) and (b) correspond to the cases of equilateral and isosceles triangles, respectively. a) The parameters in this subfigure are $L=L'$, $\varphi_{1}=-\varphi_{2}=\varphi,\ \varphi_{3}=0$, $t'=t=0.8\Delta_0,\ \epsilon'=\epsilon=0.55\Delta_0$. The two distinct values of the spin-orbit phase $\varphi_{SO}'=\varphi_{SO}=1.6$ and $\varphi_{SO}'=\varphi_{SO}=2.4$ are considered. b) The parameters in this subfigure are $L=0.3L'$, $\varphi_{1}=-\varphi_{2}=\varphi,\ \varphi_{3}=0$, $t=10t'/3=0.7\Delta_0,\ \epsilon=10\epsilon'/3=0.35\Delta_0$. The two distinct values of the spin-orbit phases $\varphi_{SO}=0.3\varphi_{SO}'=2.2,\ 2.6$ are considered.}
               \label{fig:matrix_elements}
\end{figure}

As explained in the main text, the conservation of the pseudospin quantum number inhibits transitions between the lowest pair of spin-split Andreev bound states. In this regime, the system can no longer function as an ASQ and can only be manipulated within the even parity sector as a conventional Andreev qubit. This involves breaking the Cooper pairs in the even parity ground state to populate pairs of neighboring excitations on top of the Fermi sea. In Figure \ref{fig:matrix_elements}, we detail this process along with the corresponding matrix elements.

To couple pseudospins with supercurrents, time-reversal symmetry must be broken by additional perturbations. In this section, we analyze the impact of weak external magnetic fields, assuming that their orbital effects are negligible and that the primary influence on the electrons is described by the Zeeman interaction.

\subsection{Effective Zeeman Hamiltonian}
Applying the in-plane magnetic field $\vec{B} = B (\cos \vartheta \, \vec{e}_x + \sin \vartheta \vec{e}_y)$, we find perturbative corrections to the wire Green's functions of Eqs. \eqref{GF_W}, \eqref{GF_W_32}, \eqref{GF_W_13} due to Zeeman term $\frac{g \mu_B B}{2} (\cos \vartheta \, \sigma_y - \sin \vartheta \, \sigma_x)$ (recall that this form emerges due to the initial global rotation of the Pauli matrices). In particular,
\begin{align}
    G^{(21)}_{W,B} (x,x') &\approx G^{(21)}_{W} (x,x') - \frac{g \mu_B B \sin \vartheta}{2} \mathcal{I}_{\parallel} (x,x') e^{-i \sigma_x k_R (x-x')} \sigma_x + \frac{g \mu_B B \cos \vartheta}{2} \mathcal{I}_{\perp}  (x,x') e^{-i \sigma_x k_R (x+x'-L) }  \sigma_y ,
\end{align}
where
\begin{align}
    \mathcal{I}_{\parallel} (x,x') &= \int_0^L d x''  G^{(21)}_{W,0} (x,x'')  G^{(21)}_{W,0} (x'',x'), \\
    \mathcal{I}_{\perp}  (x,x') &= \int_0^L d x''  G^{(21)}_{W,0} (x,x'') G^{(21)}_{W,0} (x'',x')  e^{ 2i \sigma_x k_R (x'' -L/2)}.
\end{align}
Analogously we evaluate
\begin{align}
    G^{(32)}_{W,B} (x,x') &\approx G^{(32)}_{W} (x,x') \nonumber \\
    &-\frac{g \mu_B B \sin (\beta  - \vartheta )}{2} U \mathcal{I}'_{\parallel} (x,x')   e^{-i \sigma_x k_R (x-x')}    \sigma_x   U^{\dagger} -\frac{g \mu_B B\cos (\beta  - \vartheta )}{2} U \mathcal{I}'_{\perp} (x,x')  e^{-i \sigma_x k_R (x+x' - L')}   \sigma_y U^{\dagger}, \\
    G^{(13)}_{W,B} (x,x') &\approx G^{(32)}_{W} (x,x') \nonumber \\
    &+\frac{g \mu_B B \sin ( \beta + \vartheta )}{2} U^{\dagger} \mathcal{I}'_{\parallel} (x,x')  e^{-i \sigma_x k_R (x-x')}  \sigma_x U -\frac{g \mu_B B\cos ( \beta  + \vartheta )}{2} U^{\dagger} \mathcal{I}'_{\perp} (x,x')  e^{-i \sigma_x k_R (x+x' - L')}  \sigma_y  U ,
\end{align}
where $\mathcal{I}'_{\parallel} (x,x')$ and $\mathcal{I}'_{\perp} (x,x')$ are obtained from $\mathcal{I}_{\parallel} (x,x')$ and $\mathcal{I}_{\perp} (x,x')$ by replacing $L \to L'$.

In turn, the out-of-plane magnetic field $\vec{B} = B \vec{e}_z$ generates the corrections
\begin{align}
    G^{(21)}_{W,B} (x,x') &\approx G^{(21)}_{W} (x,x') +  \frac{g \mu_B B}{2} \mathcal{I}_{\perp}  (x,x') e^{-i \sigma_x k_R (x+x' -L) }  \sigma_z , \\
    G^{(32)}_{W,B} (x,x') &\approx G^{(32)}_{W} (x,x') +  \frac{g \mu_B B}{2} U \mathcal{I}'_{\perp}  (x,x') e^{-i \sigma_x k_R (x+x'-L') }  \sigma_z U^{\dagger}, \\
     G^{(13)}_{W,B} (x,x') &\approx G^{(13)}_{W} (x,x') +  \frac{g \mu_B B}{2} U^{\dagger} \mathcal{I}'_{\perp}  (x,x') e^{-i \sigma_x k_R (x+x' -L') }  \sigma_z U.
\end{align}

\begin{figure}[h!]
\centering
      \includegraphics[width=0.95\textwidth]{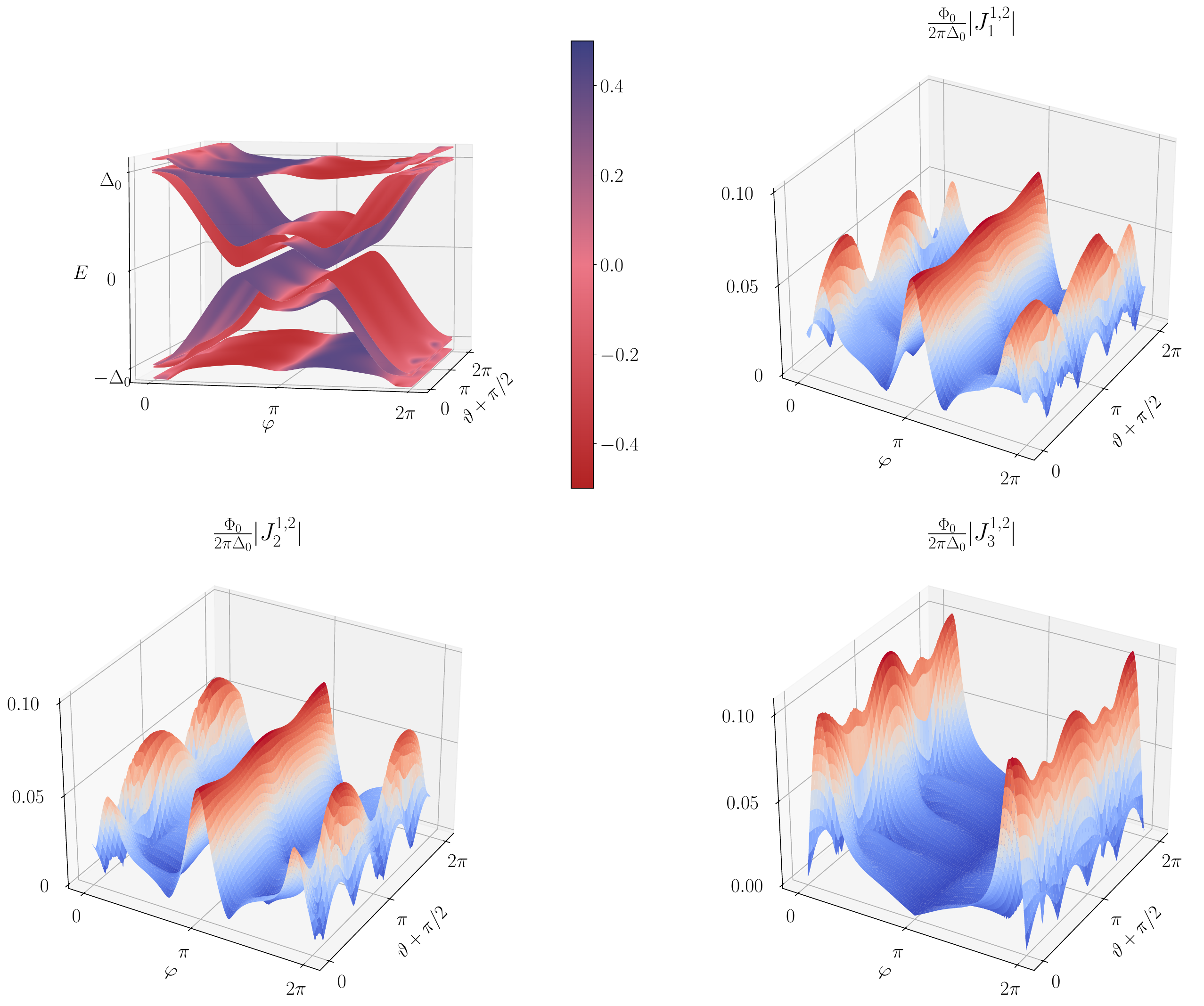}
                \caption{Numerical data for an equilateral triangle ($L = L'$) in the in-plane magnetic field $\vec{B} = B(\cos\vartheta\vec{e}_{x}+\sin\vartheta\vec{e}_{y})$ with parameters $\epsilon = 0.32\Delta_0$, $t = 0.70\Delta_0$, $\varphi_{SO} = 2.6$, and $\frac{ g \mu_B B L}{2 v_{F,S} \sin^2 (k_F L)} = 0.1$. The top left panel illustrates the flow of the energy-phase relation as the  field's orientation changes, with the states color-coded based on their pseudo-spin expectation values. The other panels show the phase $\varphi$ and the field's angle $\vartheta$ dependence of the matrix elements between the lowest pair of pseudo-spin-split Andreev bound states (with the color-coding additionally emphasizing magnitudes of their absolute values).}
               \label{fig:trnasitions_mag_parallel}
\end{figure}

Further, we consider
\begin{align}
    -\frac{1}{8 m_W^2} \lim_{x \to 0^+} \partial_x^2 \mathcal{I}_{\parallel} (x,x)&= -\frac{1}{8 m_W^2} \lim_{x \to L^-} \partial_x^2 \mathcal{I}_{\parallel} (x,x)= - \frac{L}{2 \sin^2 (k_F L)}, \\
     \frac{1}{4 m_W^2} \lim_{x \to 0^+} \lim_{x' \to L^-} \partial_x \partial_{x'} \left[ \mathcal{I}_{\parallel} (x,x') e^{-i \sigma_x k_R (x-x')} \right] &= e^{i \sigma_x k_R L} \frac{L}{2 \sin^2 (k_F L)} \cos (k_F L), \\
     \frac{1}{4 m_W^2} \lim_{x \to L^-} \lim_{x' \to 0^+} \partial_x \partial_{x'} \left[ \mathcal{I}_{\parallel} (x,x') e^{-i \sigma_x k_R (x-x')} \right] &= e^{-i \sigma_x k_R L} \frac{L}{2 \sin^2 (k_F L)} \cos (k_F L);
\end{align}
\begin{align}
    -\frac{1}{8 m_W^2} \lim_{x \to 0^+} \partial_x^2 \left[ \mathcal{I}_{\perp} (x,x) e^{-i \sigma_x k_R (2 x -L)} \right] &= - e^{i \sigma_x k_R L}  \frac{ L}{2 \sin^2 (k_F L)} \frac{\sin (k_R L)}{k_R L}, \\
    -\frac{1}{8 m_W^2} \lim_{x \to L^-} \partial_x^2 \left[ \mathcal{I}_{\perp} (x,x) e^{-i \sigma_x k_R (2 x -L)} \right] &= - e^{-i \sigma_x k_R L} \frac{ L}{2 \sin^2 (k_F L)} \frac{\sin (k_R L)}{k_R L}, \\
    \frac{1}{4 m_W^2} \lim_{x \to 0^+} \lim_{x' \to L^-} \partial_x \partial_{x'} \left[ \mathcal{I}_{\perp} (x,x') e^{-i \sigma_x k_R (x+x' -L)} \right] &=  \frac{L}{2 \sin^2 (k_F L)} \cos (k_F L) \frac{\sin (k_R L)}{k_R L}, \\
    \frac{1}{4 m_W^2} \lim_{x \to L^-} \lim_{x' \to 0^+} \partial_x \partial_{x'} \left[ \mathcal{I}_{\perp} (x,x') e^{-i \sigma_x k_R (x+x'-L)} \right] &= \frac{L}{2 \sin^2 (k_F L)} \cos (k_F L) \frac{\sin (k_R L)}{k_R L}.
\end{align}

Following the general recipes of Ref. \onlinecite{Piasotski2024} and using the above expressions, we establish an additional contribution $b_{\text{eff}}$ to the effective Hamiltonian $h_{\text{eff}}$, which is induced by the Zeeman interaction and which is not spanned by the matrix $\tau_z$.
 For the in-plane magnetic field, the matrix elements of $b_{\text{eff}}$ in the contacts' basis are expressed by
\begin{align}
    b_{11} &= -b\sin \vartheta  \sigma_x + b \cos \vartheta   \frac{\sin (k_R L)}{k_R L} e^{i \sigma_x k_R L} \sigma_y +b' \sin (\beta +\vartheta ) U^{\dagger} \sigma_x U - b' \cos (\beta + \vartheta )   \frac{\sin (k_R L')}{k_R L'} U^{\dagger} e^{-i \sigma_x k_R L'} \sigma_y  U , \\
    b_{22} &= -b'\sin (\beta - \vartheta )  U \sigma_x U^{\dagger} - b' \cos (\beta - \vartheta )  \frac{\sin (k_R L')}{k_R L'} U e^{i \sigma_x k_R L'} \sigma_y U^{\dagger}-b\sin \vartheta  \sigma_x + b \cos \vartheta   \frac{\sin (k_R L)}{k_R L} e^{-i \sigma_x k_R L} \sigma_y  , \\
    b_{33} &= b' \sin (\beta +\vartheta ) U^{\dagger} \sigma_x U - b' \cos (\beta + \vartheta )   \frac{\sin (k_R L')}{k_R L'} U^{\dagger} e^{i \sigma_x k_R L'} \sigma_y  U \nonumber \\
    &-b'\sin (\beta - \vartheta )  U \sigma_x U^{\dagger} - b' \cos (\beta - \vartheta )  \frac{\sin (k_R L')}{k_R L'} U e^{-i \sigma_x k_R L'} \sigma_y U^{\dagger} , \\
    b_{12} &= (b_{21})^{\dagger} = b \cos( k_F L ) \sin \vartheta  e^{i \sigma_x k_R L}  \sigma_x -  b \cos( k_F L ) \cos \vartheta  \frac{\sin (k_R L)}{k_R L} \sigma_y, \\
    b_{23} &= (b_{32})^{\dagger} =  b' \cos( k_F L' )\sin (\beta -\vartheta ) U e^{i \sigma_x k_R L'}  \sigma_x U^{\dagger} +  b' \cos( k_F L' ) \cos (\beta - \vartheta ) \frac{\sin (k_R L')}{k_R L'} U \sigma_y U^{\dagger}, \\
    b_{31} &=(b_{13})^{\dagger} = - b' \cos( k_F L' ) \sin ( \beta + \vartheta ) U^{\dagger} e^{i \sigma_x k_R L'}  \sigma_x U +  b' \cos( k_F L' ) \cos ( \beta + \vartheta ) \frac{\sin (k_R L')}{k_R L'} U^{\dagger} \sigma_y U,  
\end{align}
where
\begin{align}
    b = \frac{ \Delta_0 g \mu_B B L}{2 v_{F,S} \sin^2 (k_F L)}, \quad  b' = \frac{ \Delta_0 g \mu_B B L'}{2 v_{F,S} \sin^2 (k_F L')},
\end{align}
while for the out-of-plane magnetic field the matrix elements of $b_{\text{eff}}$ are expressed by
\begin{align}
    b_{11} &=  b   \frac{\sin (k_R L)}{k_R L} e^{i \sigma_x k_R L} \sigma_z  + b'    \frac{\sin (k_R L')}{k_R L'} U^{\dagger} e^{-i \sigma_x k_R L'} \sigma_z  U , \\
    b_{22} &=  b'   \frac{\sin (k_R L')}{k_R L'} U e^{i \sigma_x k_R L'} \sigma_z U^{\dagger} + b  \frac{\sin (k_R L)}{k_R L} e^{-i \sigma_x k_R L} \sigma_z  , \\
    b_{33} &= b'   \frac{\sin (k_R L')}{k_R L'} U^{\dagger} e^{i \sigma_x k_R L'} \sigma_z U + b' \frac{\sin (k_R L')}{k_R L'} U e^{-i \sigma_x k_R L'} \sigma_z U^{\dagger} ,\\
    b_{12} &= (b_{21})^{\dagger} = -  b \cos( k_F L )  \frac{\sin (k_R L)}{k_R L} \sigma_z, \\
    b_{23} &= (b_{32})^{\dagger} = -  b' \cos( k_F L' )  \frac{\sin (k_R L')}{k_R L'}  \sigma_z , \\
    b_{31} &=(b_{13})^{\dagger} = -  b' \cos( k_F L' )  \frac{\sin (k_R L')}{k_R L'} \sigma_z .
\end{align}

\begin{figure}[h!]
\centering
      \includegraphics[width=0.95\textwidth]{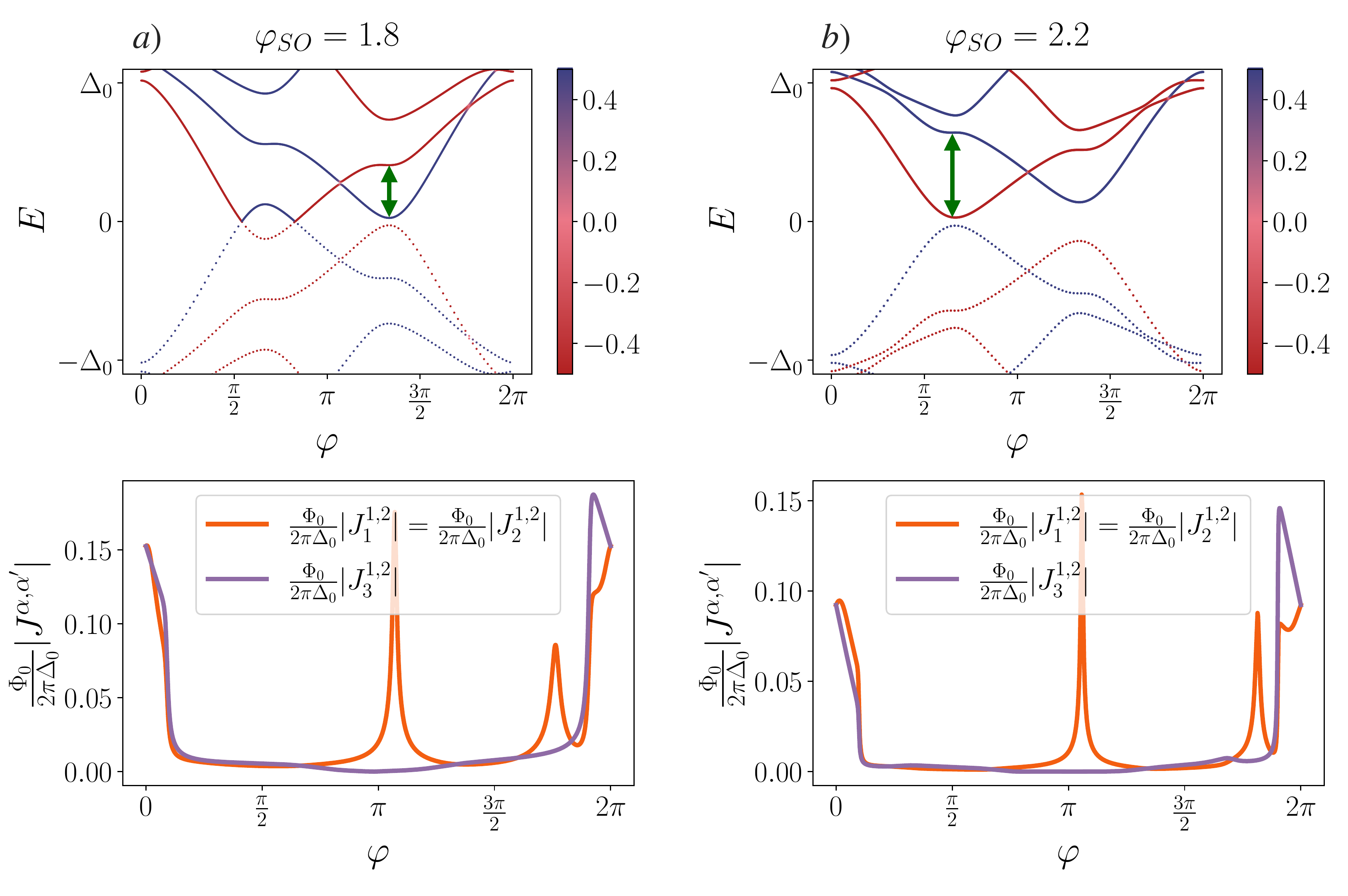}
                \caption{Numerical data for an equilateral triangle ($L=L'$) in the out-of-plane  magnetic field $\vec{B}=B\vec{e}_{z}$ with $\epsilon=0.65\Delta_0,\ t=0.75\Delta_0$ and $\frac{ g \mu_B B L}{2 v_{F,S} \sin^2 (k_F L)}=0.1$. The top panels show the energy phase relation of the junction for two values of spin-orbit angle: a) $\varphi_{SO}=1.8$ and b) $\varphi_{SO}=2.2$. The color gradient shows the pseudo-spin expectation values, while dotted and solid lines are used to differentiate between hole and particle-like states. The bottom panels show the phase-dispersion of the current matrix elements between the lowest pair of pseudo-spin-split Andreev bound states, previously (at $B=0$) being impossible by the virtue of pseudo-spin conservation.}
               \label{fig:trnasitions_mag_perp}
\end{figure}

\subsection{Discussion}
We illustrate the effects of in-plane and out-of-plane magnetic fields in Figs. \ref{fig:trnasitions_mag_parallel} and \ref{fig:trnasitions_mag_perp}, respectively. To facilitate data analysis, we color the energy levels in both figures based on the expectation value of the pseudo-spin. The colormap ranges from blue for pseudo-spin-up particles to red for pseudo-spin-down particles, with pink representing a $50-50$ admixture (pseudo-spin-zero).

As one may anticipate from the plots, the primary effect of the in-plane magnetic field is to lift the degeneracy at the time-reversal-invariant phases $\varphi=0\mod\pi$, while also breaking the reflection symmetry of the spectrum about these phase values. As one may infer from Fig. \ref{fig:trnasitions_mag_parallel}, the pseudospin projection of the Andreev levels approaches zero at the anticrossing points, which aligns with the intuitive expectation. As can be deduced from the accompanying panels, this effect has a clear consequence on the matrix elements of the current operator between the lowest pair of pseudospin-split Andreev bound states, amplifying them around the time-reversal-invariant phases. 

We report that the matrix elements $J_{j}^{1, -2}$, which are inherently present in the system, maintain their order of magnitude. The primary effect of the Zeeman field is to break their reflection symmetry and induce an asymmetry between the matrix elements of the current at the triangle's vertices $j = 1$  and $j = 2$. 

In Fig. \ref{fig:trnasitions_mag_perp}, we present data for a sample subjected to an out-of-plane magnetic field. In this example, as well as in others we have studied, we observe significantly stronger asymmetry effects in the system's spectrum due to the Zeeman interaction. We note, however, that the gaps at the avoided crossings exhibit comparable orders of magnitude for both in-plane and out-of-plane fields, leading to a similar enhancement in the transition matrix elements between the lowest pair of pseudo-spin-split bound states.

\section{Interaction between two qubits}
\label{ap: Interaction}
\begin{figure}[h!]
\centering
                \includegraphics[width=0.9\textwidth]{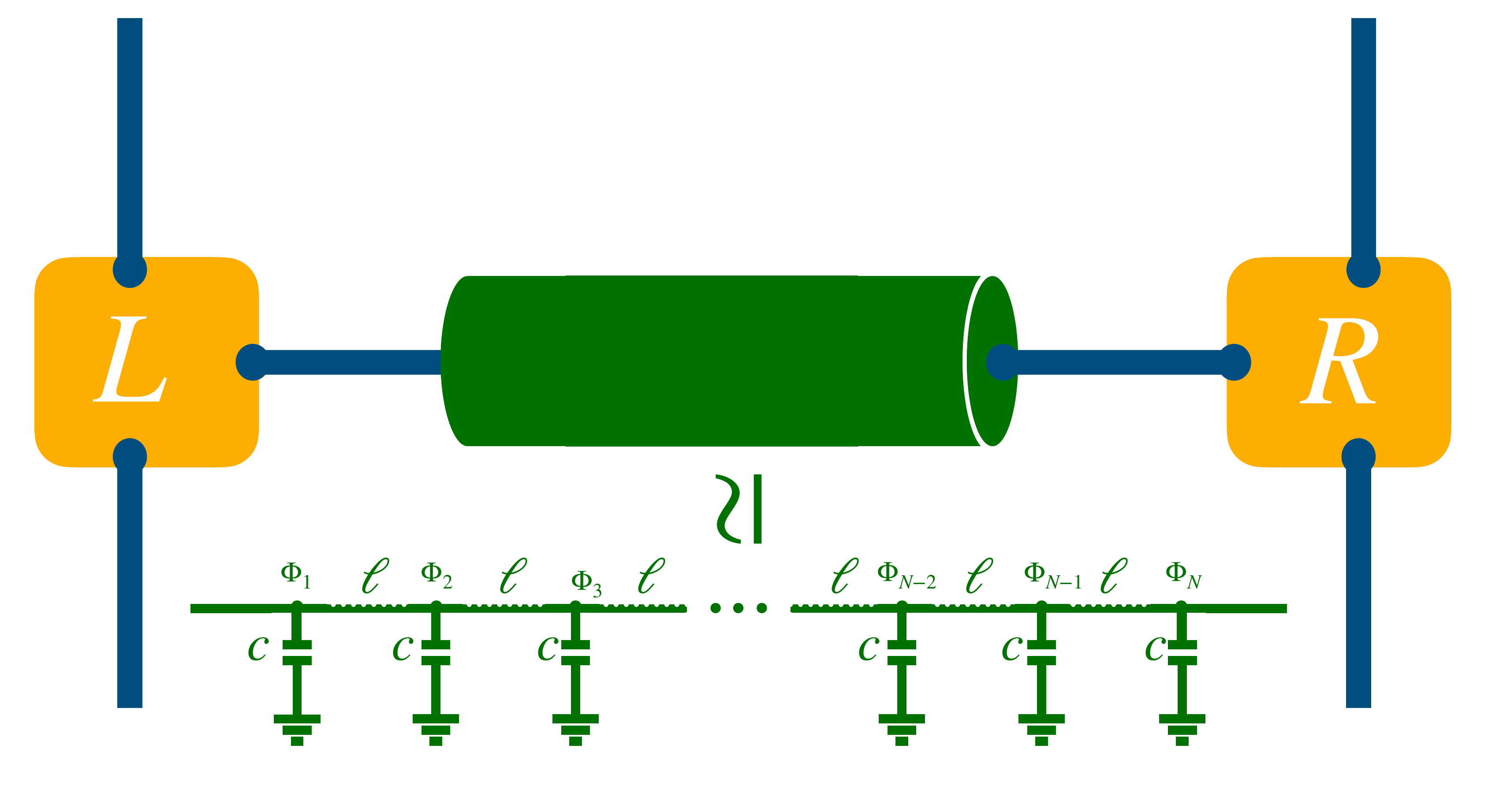}
                \caption{A schematic representation of the two qubit coupling through a superconducting filament, which is modeled by a transmission line with lumped elements.}
               \label{fig:Coupling}
\end{figure}

\subsection{General consideration}
Consider a pair of the TASQs connected by a superconducting filament as illustrated in Fig. \ref{fig:Coupling}. If the filament's length $L_{TL}$ is sufficiently large, $L_{TL} \gg \xi$, the interaction between the qubits mediated by the Cooper pair tunneling can be ignored, such that the Hamiltonian of the system is given by
\begin{align}
    \widehat{H}^{(0)}_{q}=\frac{1}{2}\widehat{\Psi}_{L}^{\dagger}h_{\text{eff}, L}\widehat{\Psi}_{L}+\frac{1}{2}\widehat{\Psi}_{R}^{\dagger}h_{\text{eff}, R}\widehat{\Psi}_{R}.
\end{align}
Here, $\widehat{\Psi}_{L}$ and $\widehat{\Psi}_{R}$ represent the extended Nambu spinors associated with the low-energy excitations of left and right islands, respectively. Indeed, since the superconducting excitations are gapped, the resulting on-shell interaction mediated by these excitations, $\frac{1}{2} \widehat{\Psi}_{R}^{\dagger} v_{RL} \widehat{\Psi}_{L} + \text{h.c.}$, is exponentially suppressed, $||v_{RL}|| \sim e^{-L_{TL} / \xi}$.

A significant interaction effect arises from the fluctuations of the electromagnetic field in the filament. To illustrate this, we assume that both qubits are coupled to it at their third ports, and examine the quantized flux fluctuations  $\varphi_{3, S} \rightarrow \varphi_{3, S} + \frac{2\pi \delta\widehat{\Phi}^{(S)}}{\Phi_0} $. Here, $\delta\widehat{\Phi}^{(S)}$ represents the fluctuating field at the filament's ends $S=L,R$, which is treated as a bosonic degree of freedom  (see below for a specific model). With this in mind, the total Hamiltonian takes the form $\widehat{H}\simeq \widehat{H}^{(0)}_{q}+\widehat{H}^{(0)}_{\Phi}+\widehat{V}$, where
\begin{align}
\widehat{H}^{(0)}_{\Phi}=&\sumint_{s}\omega_{s}\hat{a}_{s}^{\dagger}\hat{a}_{s},\quad \widehat{V}=\sum_{S=L,R}\widehat{J}^{(S)}_{3}\delta\widehat{\Phi}^{(S)}, \quad \delta\widehat{\Phi}^{(S)}=\sumint_{s}(g^{(S)}_{s}\hat{a}_{s}+g_{s}^{(S)*}\hat{a}_{s}^{\dagger}), \label{radiation_ham}\\
    \widehat{J}^{(S)}_{3}=&\frac{1}{2}\widehat{\Psi}^{\dagger}_{S}J_{3}^{(S)}\widehat{\Psi}_{S}, \quad J_{3}^{(S)}=\frac{2\pi}{\Phi_{0}}\frac{\partial h_{\text{eff}, S}}{\partial \varphi_{3, S}}=\frac{i\pi}{\Phi_{0}}\left[\tau_{z}\ket{3}\bra{3}, h_{\text{eff}, S}\right].
\end{align}
Expanding the quasiparticle field operators in the eigenbasis
\begin{align}
    \widehat{\Psi}_{S}=\sum_{\alpha}\psi_{\alpha, S}\hat{\gamma}_{\alpha, S}, \quad \hat{\gamma}_{-\alpha, S}=\hat{\gamma}_{\alpha, S}^{\dagger},\ \{\hat{\gamma}_{\alpha, S}, \hat{\gamma}_{\alpha', S}\}=\delta_{S, S'}\delta_{\alpha, -\alpha'}, \quad \{\hat{\gamma}_{\alpha, S}, \hat{\gamma}_{\alpha', S'}^{\dagger}\}=\delta_{S, S'}\delta_{\alpha, \alpha'},
\end{align}
we obtain
\begin{align}
    \widehat{H}_{q}^{(0)}=\frac{1}{2}\sum_{S=L,R}\sum_{\alpha}E_{\alpha}^{(S)}\hat{\gamma}_{\alpha, S}^{\dagger}\hat{\gamma}_{\alpha, S}, \quad \widehat{J}^{(S)}_{3}=\frac{1}{2}\sum_{\alpha, \alpha'}J_{3, S}^{\alpha, \alpha'}\hat{\gamma}_{\alpha, S}^{\dagger}\hat{\gamma}_{\alpha', S}.
\end{align}
Performing the Schrieffer–Wolff transformation $\widehat{H}'=e^{\widehat{W}}\widehat{H}e^{-\widehat{W}}$ with the generator 
\begin{align}
    \widehat{W}=&-\frac{1}{2}\sum_{S=L,R}\sum_{\alpha, \alpha'}\sumint_{s}J_{3, S}^{\alpha, \alpha'}\hat{\gamma}_{\alpha, S}^{\dagger}\hat{\gamma}_{\alpha', S}\left[\frac{g_{s}^{(S)}}{E_{\alpha'}^{(S)}-E_{\alpha}^{(S)}+\omega_{s}}\hat{a}_{s}+\frac{g_{s}^{(S)*}}{E_{\alpha'}^{(S)}-E_{\alpha}^{(S)}-\omega_{s}}\hat{a}_{s}^{\dagger}\right],
\end{align}
we derive an effective current-current interaction 
\begin{align}
    \widehat{V}_{\text{eff}}=-\frac{1}{8}\sum_{S=L,R}\sum_{S'=L,R}\sum_{\alpha, \alpha'}\sum_{\beta, \beta'}\sumint_{s}J_{3, S}^{\alpha, \alpha'}J_{3, S'}^{\beta, \beta'}\left(\frac{g_{s}^{(S)}g_{s}^{(S')*}}{\omega_{s}+(E_{\alpha'}^{(S)}-E_{\alpha}^{(S)})}+\frac{g_{s}^{(S)*}g_{s}^{(S')}}{\omega_{s}-(E_{\alpha'}^{(S)}-E_{\alpha}^{(S)})}\right)\hat{\gamma}_{\alpha, S}^{\dagger}\hat{\gamma}_{\alpha', S}\hat{\gamma}_{\beta, S'}^{\dagger}\hat{\gamma}_{\beta', S'} \, .
    \label{V_eff}
\end{align}

Next, we define the computational basis for the two qubits corresponding to an odd ground state in each of them:
\begin{align}
    \ket{g, g}=\ket{g, L}\otimes\ket{g, R},\quad \ket{g, e}=\ket{g, L}\otimes\ket{e, R}, \quad \ket{e, g}=\ket{e, L}\otimes\ket{g, R}, \quad \ket{e, e}=\ket{e, L}\otimes\ket{e, R} , \label{two_qubit_states_odd}
\end{align}
where
\begin{align}
    \ket{g, S}=\hat{\gamma}_{1, S}^{\dagger}\prod_{\alpha>0}\hat{\gamma}_{\alpha, S}\ket{0}, \quad \ket{e, S}=\hat{\gamma}_{2, S}^{\dagger}\hat{\gamma}_{1, S}\ket{g, S},
\end{align}
and the Pauli operators are defined according to
\begin{align}
    \sigma_{+}^{(S)}=\ket{e, S}\bra{g, S}, \quad \sigma_{-}^{(S)}=\left(\sigma_{+}^{(S)}\right)^{\dagger}, \quad \sigma_{z}^{(S)}=\ket{e, S}\bra{e, S}-\ket{g, S}\bra{g, S}.
\end{align}

Projecting the effective interaction \eqref{V_eff} and the bare Hamiltonian $\widehat{H}_q^{(0)}$ onto the four-dimensional subspace spanned by the states \eqref{two_qubit_states_odd}, we reveal
\begin{align}
    H_{\text{eff}}\simeq \frac{1}{2}\sum_{S=L,R}\mathcal{E}_{S}\sigma_{z}^{(S)}-\mathcal{J}_{z}\sigma_{z}^{(R)}\sigma_{z}^{(L)}-\left(\mathcal{J}_{1}\sigma_{+}^{(R)}\sigma_{+}^{(L)}+\mathcal{J}_{1}^{*}\sigma_{-}^{(R)}\sigma_{-}^{(L)}+\mathcal{J}_{0}\sigma_{+}^{(R)}\sigma_{-}^{(L)}+\mathcal{J}_{0}^{*}\sigma_{+}^{(L)}\sigma_{-}^{(R)}\right)+\tilde{E}^{(0)}, \label{two-qubit_hamiltonian}
\end{align}
where
\begin{align}
    \mathcal{E}_{S}=&\mathcal{E}_{S}^{(0)}-\sum_{S'=L,R}\frac{J_{3, S'}^{2, 2}+J_{3, S'}^{1, 1}}{2}\frac{J_{3, S}^{2, 2}-J_{3, S}^{1, 1}}{2}\sumint_{s}\frac{\text{Re}\{g_{s}^{(S)}g_{s}^{(S')*}\}}{\omega_{s}},\quad \mathcal{E}_{S}^{(0)}=E_{2}^{(S)}-E_{1}^{(S)}, \\
    \mathcal{J}_{z}=&\frac{1}{2}\frac{J_{3, R}^{2, 2}-J_{3, R}^{1, 1}}{2}\frac{J_{3, L}^{2, 2}-J_{3, L}^{1, 1}}{2}\sumint_{s}\frac{\text{Re}\{g_{s}^{(R)}g_{s}^{(L)*}\}}{\omega_{s}},\\
    \mathcal{J}_{0}=&\frac{1}{2}J_{3, R}^{2, 1}J_{3, L}^{1, 2}\sumint_{s}\Bigg(\frac{g_{s}^{(R)}g_{s}^{(L)*}}{\omega_{s}-\mathcal{E}_{R}^{(0)}}+\frac{g_{s}^{(R)*}g_{s}^{(L)}}{\omega_{s}+\mathcal{E}_{R}^{(0)}}+\frac{g_{s}^{(L)}g_{s}^{(R)*}}{\omega_{s}+\mathcal{E}_{L}^{(0)}}+\frac{g_{s}^{(L)*}g_{s}^{(R)}}{\omega_{s}-\mathcal{E}_{L}^{(0)}}\Bigg),\\
    \mathcal{J}_{1}=&\frac{1}{2}J_{3, R}^{2, 1}J_{3, L}^{2, 1}\sumint_{s}\Bigg(\frac{g_{s}^{(R)}g_{s}^{(L)*}}{\omega_{s}-\mathcal{E}_{R}^{(0)}}+\frac{g_{s}^{(R)*}g_{s}^{(L)}}{\omega_{s}+\mathcal{E}_{R}^{(0)}}+\frac{g_{s}^{(L)}g_{s}^{(R)*}}{\omega_{s}-\mathcal{E}_{L}^{(0)}}+\frac{g_{s}^{(L)*}g_{s}^{(R)}}{\omega_{s}+\mathcal{E}_{L}^{(0)}}\Bigg),\\
    \tilde{E}^{(0)}=&E_{1}^{(S)}-\frac{1}{2}\sum_{S=L,R}\sum_{\alpha>0}E_{\alpha}^{(S)}-\sum_{S=L,R}|J_{3, S}^{2,1}|^{2}\sumint_{s}\frac{|g_{s}^{(S)}|^{2}\omega_{s}}{\omega_{s}^{2}-\mathcal{E}_{S}^{(0)2}}+\braket{g, g|\widehat{V}_{\text{eff}}|g, g}.
\end{align}

Note that the Hamiltonian \eqref{two-qubit_hamiltonian} falls into the direct sum of operators $H_{\text{eff},0}$ and $H_{\text{eff},1}$, fulfilling $[H_{\text{eff},0}, H_{\text{eff},1}]=0$ and acting on the subspaces spanned by the states $\{\ket{e, g},\ \ket{g, e} \}$ and $\{\ket{g, g},\ \ket{e, e}\}$, respectively. In the explicit expression,
\begin{align}
    H_{\text{eff},0} &= \frac{\mathcal{E}_{R}-\mathcal{E}_{L}}{2}(\ket{g, e}\bra{g, e}-\ket{e, g}\bra{e, g})-\left(\mathcal{J}_{0}\ket{e, g}\bra{g, e}+\mathcal{J}_{0}^{*}\ket{g, e}\bra{e, g}\right)+\left(\tilde{E}^{(0)}+\mathcal{J}_{z}\right)(\ket{g, e}\bra{g, e}+\ket{e, g}\bra{e, g}), \\
    H_{\text{eff},1} &= \frac{\mathcal{E}_{R}+\mathcal{E}_{L}}{2}(\ket{e, e}\bra{e, e}-\ket{g, g}\bra{g, g})-\left(\mathcal{J}_{1}\ket{e, e}\bra{g, g}+\mathcal{J}_{1}^{*}\ket{g, g}\bra{e, e}\right)+\left(\tilde{E}^{(0)}-\mathcal{J}_{z}\right)(\ket{e, e}\bra{e, e}+\ket{g, g}\bra{g, g}).
\end{align}
To couple the two subspaces, one further considers driving the system with an external time-dependent flux $\delta{\Phi}_{\text{ext}}^{(S)}(t)$. One finds the following addition to Eq. \eqref{two-qubit_hamiltonian}:
\begin{align}
    H_{\text{drive}}(t)\simeq \sum_{j}\sum_{S=L,R}\left(J_{j, S}^{1, 2}\sigma_{-}^{(S)}+J_{j, S}^{2, 1}\sigma_{+}^{(S)}+\frac{J_{j, S}^{2, 2}-J_{j, S}^{1, 1}}{2}\sigma_{z}^{(S)}+\underbrace{J_{j, S}^{1, 1}-\frac{1}{2}\sum_{\alpha>0}J_{j, S}^{\alpha, \alpha}}_{\text{ground state current}}\right)\delta{\Phi}_{\text{ext}}^{(S, j)}(t).
\end{align}
This Hamiltonian gives rise to single-qubit transitions, governed by the generators $\sigma_{\pm}^{(S)}$. Combined with the two-qubit interaction discussed earlier, it enables the implementation of arbitrary two-qubit gates.

Let us now estimate the above-derived interaction, assuming it is mediated by the linearly-dispersing plasmonic modes \cite{mooij1985propagating}. As a simple model of these plasma excitations, we consider the Hamiltonian of a one-dimensional transmission line with $N$ internal nodes in the discrete model:
\begin{align}
    \widehat{H}^{(0)}_{\Phi}=\sum_{n=1}^{N}\frac{\widehat{Q}^{2}_{n}}{2c}+\sum_{n=1}^{N}\frac{\widehat{\Phi}_{n}^{2}}{\ell}-\sum_{n=1}^{N-1}\frac{\widehat{\Phi}_{n+1}\widehat{\Phi}_{n}}{\ell}, \label{TL_hamilotonian}
\end{align}
where $\ell$ and $c$ are the inductance and capacitance of the corresponding lumped elements. The quantized, canonically conjugate flux $\widehat{\Phi}_{n}$ and charge $\widehat{Q}_{n}$ variables satisfy the commutation relations $[\widehat{Q}_{n}, \widehat{Q}_{n'}]=[\widehat{\Phi}_{n}, \widehat{\Phi}_{n'}]=0,\ [\widehat{\Phi}_{n}, \widehat{Q}_{n'}]=i\delta_{n, n'}$. To specify the interaction $\widehat{V}$ with the qubits, we identify $\delta\widehat{\Phi}^{(L)}=\widehat{\Phi}_{1}$ and $\delta\widehat{\Phi}^{(R)}=\widehat{\Phi}_{N}$.

We diagonalize the Hamiltonian  \eqref{TL_hamilotonian} in the basis of standing waves 
\begin{align}
    \widehat{\Phi}_{n}=&\sum_{k=1}^{N}\frac{\sin\frac{\pi k n}{N+1}}{\sqrt{c\omega_{k}(N+1)}}(\hat{a}_{k}+\hat{a}_{k}^{\dagger}), \label{expansion} \\
    \widehat{Q}_{n}=&-i\sum_{k=1}^{N}\sqrt{\frac{c\omega_{k}}{N+1}}\sin\frac{\pi k n}{N+1}(\hat{a}_{k}-\hat{a}_{k}^{\dagger}), \\
\widehat{H}^{(0)}_{\Phi}=&\sum_{k=1}^{N}\omega_{k}\left(\hat{a}_{k}^{\dagger}\hat{a}_{k}+\frac{1}{2}\right),\quad \omega_{k}=2\Omega_{p}\sin\frac{\pi k}{2(N+1)},
\end{align}
where $\Omega_{p}=\frac{1}{\sqrt{c\ell}}$ is the plasma oscillation frequency, and $\hat{a}_{k}^{\dagger},\ \hat{a}_{k}$ are the corresponding plasmon creation and annihilation operators, satisfying the bosonic commutation relations $[\hat{a}_{k}, \hat{a}_{k'}]=0$, $[\hat{a}_{k}, \hat{a}_{k'}^{\dagger}]=\delta_{k, k'}$. Note that in the continuum limit, one recovers a linearly dispersing mode 
\begin{align}
    \omega_{K}\simeq v_{p}K, \quad K=\frac{\pi k}{L_{TL}}, \quad v_{p}=\frac{\Omega_{p}L_{TL}}{N},
\end{align}
where $L_{TL}$ is the length of the transmission line.

The expansion of Eq. \eqref{expansion} helps us to identify the coupling constants
\begin{align}
    g_{k}^{(L)}=\frac{\sqrt{\ell}\Omega_{p}}{\sqrt{\omega_{k}(N+1)}}\sin\frac{\pi k}{N+1}, \quad g_{k}^{(R)}=-\frac{\sqrt{\ell}\Omega_{p}(-1)^{k}}{\sqrt{\omega_{k}(N+1)}}\sin\frac{\pi k}{N+1},
\end{align}
whose form gives the scaling presented in the main text:
\begin{align}
    \frac{g_{k}^{(S)}g_{k}^{(S')*}}{\omega_{k}}\propto \ell.
\end{align}
In particular, the Lamb shifts and the Ising coupling amount to
\begin{align}
    \mathcal{E}_{S} - \mathcal{E}_{S}^{(0)} =& -\frac{N\ell}{2(N+1)}\frac{(J_{3, S}^{2, 2})^{2}-(J_{3, S}^{1, 1})^{2}}{4}-\frac{\ell}{2(N+1)}\frac{J_{3, \bar{S}}^{2, 2}+J_{3, \bar{S}}^{1, 1}}{2}\frac{J_{3, S}^{2, 2}-J_{3, S}^{1, 1}}{2},\\
    \mathcal{J}_{z}=&\frac{\ell}{4(N+1)}\frac{J_{3, R}^{2, 2}-J_{3, R}^{1, 1}}{2}\frac{J_{3, L}^{2, 2}-J_{3, L}^{1, 1}}{2},
\end{align}
where $\bar{S}\neq S$.

\bibliography{citations_SM.bib}